\documentclass[fleqn,usenatbib,useAMS]{mnras}

\usepackage{graphicx}	
\usepackage{amsmath}	
\usepackage{amssymb}	
\usepackage{multicol}        
\usepackage{bm}		
\usepackage{pdflscape}	
\usepackage{stfloats} 
\usepackage{cite}

\usepackage[T1]{fontenc}
\usepackage{ae,aecompl}

\usepackage{newtxtext,newtxmath}

\title[Stellar abundances from low-resolution spectra]{The stellar parameters and elemental abundances from low-resolution spectra \uppercase\expandafter{\romannumeral1}: 1.2 million giants from LAMOST DR8}

\author{Zhuohan Li$^{1,2}$}
\author[Li et al.]{
Zhuohan Li$^{1,2}$,
Gang Zhao$^{1,2}$, \thanks{Contact e-mail: \href{gzhao@nao.cas.cn}{gzhao@nao.cas.cn}}
Yuqin Chen$^{1,2}$,
Xilong Liang$^{1,2}$,
Jingkun Zhao$^{1,2}$
\\
$^{1}$ CAS Key Laboratory of Optical Astronomy, National Astronomical Observatories, Chinese Academy of Sciences, Beijing 100101, People' s Republic of China
\\
$^{2}$ School of Astronomy and space Science, University of Chinese Academy of Sciences, Beijing 100049, People’s Republic of China}

\date{Last updated 2020 June 10; in original form 2013 September 5}

\pubyear{2022}

\begin{document}
\label{firstpage}
\pagerange{\pageref{firstpage}--\pageref{lastpage}}
\maketitle

\begin{abstract}
As a typical data-driven method, deep learning becomes a natural choice for analysing astronomical data nowadays. In this study, we built a deep convolutional neural network to estimate basic stellar parameters $T\rm{_{eff}}$, log \textit{g}, metallicity ([M/H] and [Fe/H]) and [$\alpha$/M] along with nine individual elemental abundances ([C/Fe], [N/Fe], [O/Fe], [Mg/Fe], [Al/Fe], [Si/Fe], [Ca/Fe], [Mn/Fe], [Ni/Fe]). The neural network is trained using common stars between the APOGEE survey and the LAMOST survey. We used low-resolution spectra from LAMOST survey as input, and measurements from APOGEE as labels. For stellar spectra with the signal-to-noise ratio in \textit{g} band larger than 10 in the test set, the mean absolute error (MAE) is 29 K for $T\rm{_{eff}}$, 0.07 dex for log \textit{g}, 0.03 dex for both [Fe/H] and [M/H], and 0.02 dex for [$\alpha$/M]. The MAE of most elements is between 0.02 dex and 0.04 dex. The trained neural network was applied to 1,210,145 giants, including sub-giants, from LAMOST DR8 within the range of stellar parameters 3500 K < $T\rm{_{eff}}$ < 5500 K, 0.0 dex < log \textit{g} < 4.0 dex, -2.5 dex < [Fe/H] < 0.5 dex. The distribution of our results in the chemical spaces is highly consistent with APOGEE labels and stellar parameters show consistency with external high-resolution measurements from GALAH. The results in this study allow us to further studies based on LAMOST data and deepen our understanding of the accretion and evolution history of the Milky Way. The electronic version of the value added catalog is available at \url{http://www.lamost.org/dr8/v1.1/doc/vac}.
\end{abstract}




\begingroup
\let\clearpage\relax
\endgroup
\newpage

\section{Introduction}
	The accurate measurement of stellar atmospheric parameters and elemental abundances is of great significance to the study of Galactic evolution. With the development of technology, a number of modern spectroscopic surveys such as the RAdial Velocity Experiment (RAVE; \citealt{steinmetz2006radial}), Sloan Extension for Galactic Understanding and Exploration (SEGUE; \citealt{yanny2009segue}), LAMOST survey \citep{zhao2006stellar, zhao2012lamost, deng2012lamost, liu2013lss}, Gaia mission \citep{prusti2016gaia}, Gaia-ESO survey \citep{gilmore2012gaia}, Galactic Archaeology with HERMES (GALAH; \citealt{de2015galah}) and Apache Point Observatory Galactic Evolution Experiment (APOGEE; \citealt{majewski2017apache}) have provided massive amounts of data that propelled studies of the evolution history of the Milky Way.\par
	Since each system has its own particular formation and chemical enrichment history, stars born in different systems will follow their own chemical sequences \citep{helmi2020streams}. Large samples of stars with detailed chemical abundance information can support the studies of substructures such as Gaia-Enceladus \citep{helmi2018merger,belokurov2018co}, the Helmi streams \citep{helmi1999debris}, Sequoia \citep{myeong2019evidence}, Thamnos \citep{koppelman2019multiple} and kinematically selected moving groups (e.g., \citealt{zhao2009catalog}), but the availability of the chemical informations is the currnt limitation. This is when deep learning can come into play.\par
	In the context of the big data, with the improvement of computing power and the reduction of computing costs, deep learning methods have gradually begun to be widely used in astronomy. Deep learning allows computational models that are composed of multiple processing layers to learn representations of data with multiple levels of abstraction \citep{lecun2015deep}, it is a method with great potential in data mining. As one of the representatives of deep learning algorithms, neural networks have long been used in researches in the field of astronomy. \citet{von1994automated} have used a neural network with one hidden layer to classify stellar spectra from high-dispersion objective prism plates. \citet{bailer1997physical} trained an artificial neural network on a set of synthetic stellar spectra and used it to predict $T\rm{_{eff}}$, log \textit{g} and [M/H]. These are pioneering works on the use of neural networks in astronomy, and today's neural networks perform even better. A neural network developed by \citet{imig2022sdss} is able to derive $T\rm{_{eff}}$, log \textit{g}, [Fe/H], [$\alpha$/M] and $V\rm{_{mic}}$ from spectra in the MaNGA Stellar Library (MaStar; \citealt{yan2019sdss}), which cover a extensive range of stellar properties. \citet{ting2019payne} came up with a method named \texttt{The Payne} to estimate $T\rm{_{eff}}$, log \textit{g}, [Fe/H] and the abundances of 14 elements, after their related work \citep{ting2017measuring}. The main component of \texttt{The Payne} is a fully connected network with two hidden layers, it maps a set of stellar labels to the spectrum to enforce that the elemental abundances are derived from their corresponding absorption features instead of astrophysical correlations. Inheriting essential ingredients from \texttt{The Payne} and a data-driven method called \texttt{The Cannon} \citep{ness2015cannon,ho2016cannon}, \citet{xiang2019abundance} created the data-driven Payne (\texttt{DD-Payne}) to derive $T\rm{_{eff}}$, log \textit{g}, $V\rm{_{mic}}$ and abundances of 16 elements (C, N, O, Na, Mg, Al, Si, Ca, Ti, Cr, Mn, Fe, Co, Ni, Cu, and Ba) from LAMOST spectra. A method called \texttt{StarNet} \citep{fabbro2018application} is a deep convolutional neural network trained on spectra from APOGEE, it predicts $T\rm{_{eff}}$, log \textit{g} and [Fe/H] with similar precision as the APOGEE pipeline. The same architecture is used by \citet{zhang2019938} to obtain four stellar parameters ($T\rm{_{eff}}$, log \textit{g}, [Fe/H], and [$\alpha$/M]), carbon and nitrogen abundances from LAMOST spectra. Their network is trained with common stars between APOGEE and LAMOST, using spectra from LAMOST and stellar labels from APOGEE, such as \citet{ho2017label}. \citet{leung2019deep} designed a neural network to determine stellar parameters and abundances of 18 individual elements for high-resolution spectra from APOGEE. They also developed an open-source python package called \texttt{astroNN} which can learn from incomplete data and give the uncertainty of the predictions. \citet{liang2019elemental} built neural networks based on \texttt{astroNN} package to estimate $T\rm{_{eff}}$, log \textit{g}, [Fe/H] along with abundances of 12 individual elements for low-resolution spectra from LAMOST survey. Their neural networks are trained with common stars between APOGEE DR14 \citep{abolfathi2018fourteenth} and LAMOST DR5 \citep{luo2015first}, and they trained networks for each label independently.\par
	In this work, we adopted a deep learning approach to estimate atmospheric parameters ($T\rm{_{eff}}$, log \textit{g}, [Fe/H], [M/H], [$\alpha$/M]) and nine individual elemental abundances ([C/Fe], [N/Fe], [O/Fe], [Mg/Fe], [Al/Fe], [Si/Fe], [Ca/Fe], [Mn/Fe], [Ni/Fe]) for 1,210,145 giant stars in the newly released LAMOST DR8 catalog, which allows more possibilities for researches based on LAMOST data and would be helpful for us to understand the accretion and evolution history of the Milky Way.\par
	This paper is organized as follows: in section~\ref{sec:2}, we introduce the neural network model we used, our data treatments and training steps. Section~\ref{sec:3} presents the performance of our neural network on the training set and the test set. In section~\ref{sec:4}, we apply the trained neural network to 1,210,145 giants from LAMOST DR8 to estimate the stellar atmospheric parameters and the abundances of nine elements for these stars, and the results are compared with external high-resolution measurements. Finally, we summarize this study in section~\ref{sec:5}.

\section{Method}
\label{sec:2}
We build our deep learning model based on \texttt{astroNN}, and train it with spectra from LAMOST and stellar labels from APOGEE, which is similar to \citet{liang2019elemental}. Compared to previous data releases, there are more stars in common between the newly released data from LAMOST and APOGEE, which allows us to train a deeper neural network with more nodes and more powerful fitting ability to estimate stellar parameters and nine elemental abundances simultaneously rather than make independent predictions for each label which could break the consistncy between the labels. In addition to the increase in training samples and the improvement of the neural network structure, we took advantage of the custom loss function of \texttt{astroNN} to achieve training with defective labels. This approach, along with our weighting of metal-poor stars, makes our neural network performs better than the previous work, which is reflected in the smaller deviations of our residuals and smaller uncertainty in the predictions.\par
\subsection{The \texttt{astroNN} package}
	The \texttt{astroNN} is an open-source python package developed by \citet{leung2019deep}. We build our own neural network that can give the uncertainty of the predictions with this package. \citet{hinton2012improving} proposed a technique called dropout which is primarily used to prevent overfitting and neurons will be temporarily dropped from the network according to a certain probability during the training process. \citet{gal2016dropout} showed that a neural network with arbitrary depth and non-linearities, with dropout applied before every weight layer, is mathematically equivalent to an approximation to the probabilistic deep Gaussian process \citep{damianou2013deep}. This lays the theoretical foundation for \texttt{astroNN} to use dropout variational inference as an approximation to Bayesian neural networks. For uncertainty estimation, the \texttt{astroNN} package runs Monte Carlo dropout with its custom layer named \texttt{MCDropout} N times in forward passes through the network. The neural network has different predictions in every forward pass through the network due to the randomness of dropout process. The mean value of predictions is taken as the final prediction and the standard deviation of predictions becomes the model uncertainty. In addition to the model uncertainty, the neural network also gives a predictive uncertainty which captures noise inherent in the observations. The \texttt{astroNN} package takes the sum of model and predictive uncertainty in quadrature as the final uncertainty (see \citealt{kendall2017uncertainties}).\par
\begin{figure}
 \includegraphics[width=\columnwidth]{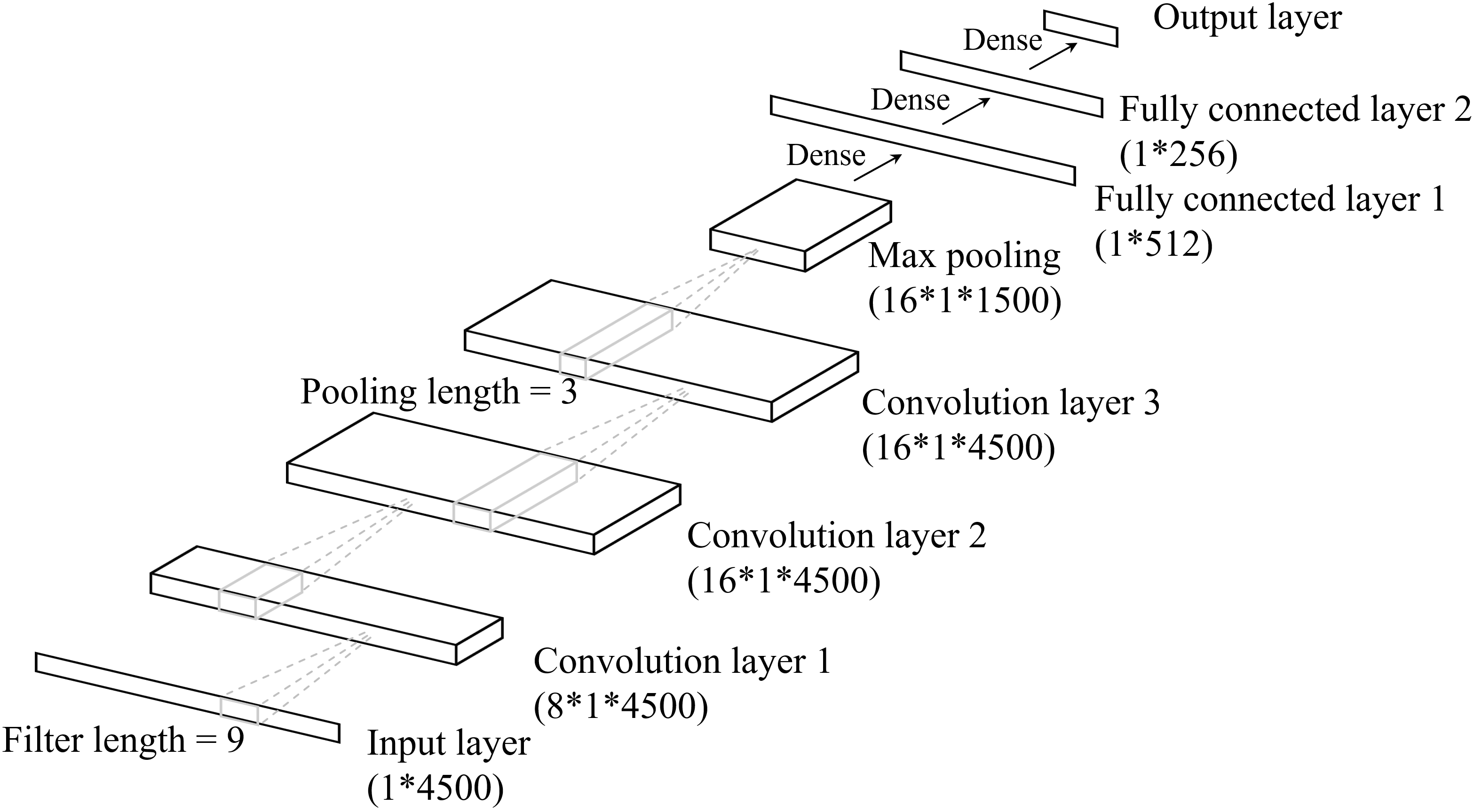}
 \caption{The structure of the convolutional neural network based on the class \texttt{BayesianCNNBase} from \texttt{astroNN}. The neural network consists of three convolutional layers and two fully connected layers, other specific parameters are marked in the figure.}
 \label{fig:Noah}
\end{figure}

	The network we used in our experiment is a deep convolutional neural network based on the class \verb'BayesianCNNBase' from \verb'astroNN'. The architecture of the network is shown in Figure~\ref{fig:Noah}. Our network structure has three convolutional layers, the third convolutional layer is followed with a max-pooling layer. The top two layers are fully connected layers. To prevent overfitting and perform dropout variational inference, a custom dropout layer from \texttt{astroNN} called \texttt{MCDropout} is added after all layers except the third convolutional layer and the second fully connected layer.

\subsection{Network training}
Our neural network was trained on spectra from LAMOST DR8 \citep{luo2015first}. The atmospherical parameters and elemental abundances from APOGEE DR17 \citep{accetta2022seventeenth} were used as labels. The Large sky Area Multi-Object	Spectroscopy Telescope (LAMOST; \citealt{cui2012large}; \citealt{zhao2012lamost}) survey is a low-resolution (R $\approx$ 1800) optical spectroscopic survey, covering the wavelength range from 3700 Å to 9000 Å, which is suitable for the search of metal-poor stars and the deriving of radial velocity and abundances for numerous scientific investigations \citep{li2015spectroscopic,liu2015preface}. With 4000 fibers placed on the focal plane, it can obtain 4000 spectra in a 5° field simultaneously, making it the telescope with the highest spectral acquisition rate in the world. It has made important contributions in searching for the first generation stars and researches of the formation and evolution history of the Milky Way. The Apache Point Observatory Galactic Evolution Experiment (APOGEE; \citealt{majewski2017apache}) is a high-resolution (R $\approx$ 22,500), high signal-to-noise ratio ($\textgreater$ 100) spectroscopic survey which observes in the wavelength range from 15100 Å to 17000 Å. The APOGEE survey makes it possible for the first time to explore the chemo-kinematical character of all Milky Way stellar subsystems using the same set of high quality data. In the seventeenth data release, both the raw spectroscopic stellar parameters and the calibrated parameters and abundances are provided. The effective temperatures are calibrated by a comparison to photometric effective temperatures and the surface gravity is calibrated using a neural network. The calibration for the elemental abundances consists of a zero point offset (see \citealt{accetta2022seventeenth} for more details). In order to obtain accurate results, only calibrated parameters and abundances were used in this study. \par
There are 7,162,019 stars with signal-to-noise ratio of the \textit{g} band (S/N\textit{g}) from LAMOST larger than 10, we cross-matched them with stars having complete $T\rm{_{eff}}$, log \textit{g}, [Fe/H], [M/H] and [$\alpha$/M] in APOGEE DR17 catalog to obtain common stars. Common stars with $T\rm{_{eff}}$ between 3500 K and 5500 K, log \textit{g} between 0.0 dex and 4.0 dex were then selected into the dataset set, with the total number of 76,220.\par
\begin{figure*}
 \includegraphics[width=\textwidth]{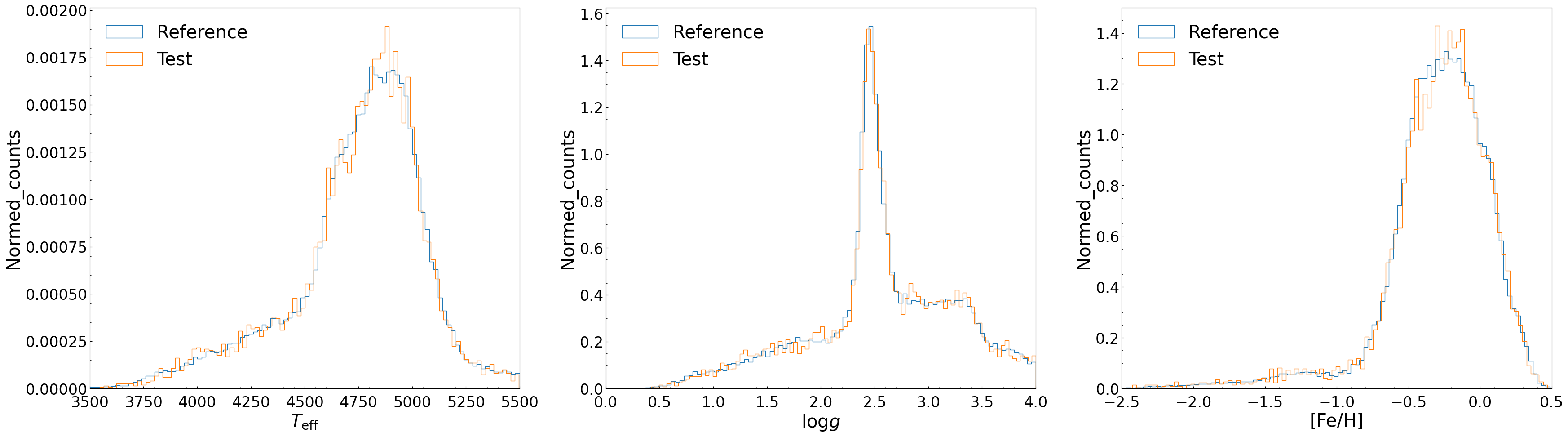}
 \caption{Distributions of the atmospherical parameters for 68,363 stars in the reference set and 7596 stars in the test set, which are shown in blue and orange, respectively. Parameters in the reference set and the test set are identically distributed.}
 \label{fig:parameter}
\end{figure*}

We intercepted the spectra from 4000 Å to 8500 Å and interpolated to ensure that each angstrom has a corresponding flux value. Following the method in \citet{ho2017label}, we performed continuum normalization by Gaussian smoothing with Gaussian kernels of width 50 Å, using the \texttt{dataset} module from \texttt{The Cannon}. During the processing above, we removed spectra for which the LAMOST Stellar Parameter Pipeline (LASP; \citealt{luo2015first}) did not give their redshift values and spectra which were heavily polluted by cosmic rays. There are 75,959 stars left in the dataset and they are then divided into reference set and test set randomly but with the same distribution in a ratio of 9:1, so that we can use as much stars as possible for training while still keeping enough samples to test the accuracy of the predictions. The reference samples were further divided into training set and cross-validation set also in the proportion of 9:1. Figure~\ref{fig:parameter} shows the parameter coverage of the reference set and the test set, a Kiel diagram will be later shown in Figure~\ref{fig:6}.
\par
	From the 7,162,019 stars, we selected 1,210,145 giants, including 181,427 sub-giants (3.5 dex < log \textit{g} < 4.0 dex), with 3500 K < $T\rm{_{eff}}$ < 5500 K, 0.0 dex < log \textit{g} < 4.0 dex, -2.5 dex < [Fe/H] < 0.5 dex according to the values given by the LASP, these stars are the target samples. After the neural network is trained, it will be used to estimate the stellar parameters and elemental abundances of the target samples from LAMOST DR8 cataloge.
\subsubsection{Labels treatment}
In general, we want every label of the sample to be complete when pre-processing the data, so we tend to discard samples with missing values. When we make predictions for multiple elements, we have to discard lots of samples, many of which have only one or a few missing values while the other values are complete. In regression tasks of the neural network, the Mean Squared Error (MSE) is the most common loss function but it can’t deal with missing values. The \texttt{astroNN} has a custom loss function, defined as
\begin{equation}
J\left(y_i,{\hat{y}}_i\right)=
\begin{cases}
\frac{1}{2}({\hat{y}}_i-y_i)^2e^{-s_i}+\frac{1}{2}(s_i) & \text{for $y_{i}$ $\neq$ NaN} \\
0 & \text{for $y_{i}$} = \text{NaN}
\end{cases}
\end{equation}
where $s_i=ln[\sigma_{known,i}^2+\sigma_{predictive,i}^2]$. Item $\sigma_{known,i}^2$ corresponds to the known uncertainty variance in the labels and $\sigma_{predictive,i}^2$ is the additional predictive variance. Defective labels will not affect the loss function, and the influence of labels with excessive uncertainty on the loss function is also weakened, so this loss function is robust enough to handle incomplete labels and high uncertainty, which makes training with defective values possible. The biggest advantage of this training method is that there is no need to remove a star from the data set because of the loss of a certain element, we can still use other abundance information that the star has. In this way, we will discard fewer samples during data pre-processing, which is particularly prominent for the metal-poor samples. Although \texttt{astroNN} can be trained with missing values, we still want the four parameters of effective temperature, surface gravity, and metallicity to be complete.\par
Due to the existence of $s_{i}$ in the loss function, the network can theoretically handle labels with large uncertainty, which will only have a small impact on the loss function. However, in order to make the results more accurate, we still set loose limits on the uncertainty of each element abundance. For $\alpha$\footnote{The total $\alpha$ abundance given by APOGEE, which is influenced by a combination of all $\alpha$ elements (O, Mg, S, Si, Ca, Ti) measured in the survey.}, C, N, O, Mg, Al, Si and Ca, the upper limits of label uncertainty are set to be 0.20 dex. Label uncertainty limits for Mn and Ni are set at 0.05 dex, as their label uncertainties are inherently small and we want the neural network predictions for them to be more accurate. Labels with uncertainty outside the range are set to be NaN and will be ignored by the loss function. These constraints can remove outliers without seriously affecting the number of non-empty labels.\par
The neural network is more inclined to fit the part where the samples are aggregated, and it is difficult for the network to fit the part where the sample distribution is sparse. Therefore, we subjectively set an effective interval for each parameter space, which excludes samples distributed in extremely sparse parts, so that the network can better fit the vast majority of the samples. After the limitation, the parameter range is still relatively wide, as shown in Table~\ref{tab:1}.\par

\begin{table}
 \caption{Effective parameter ranges of the labels. The numers in the right column represent the number of times each label appears in the dataset. Since some stars carry defective labels or some of its labels are set to be NaN during the pre-processing, the numbers vary with labels.}
 \label{tab:1}
 \begin{tabular*}{\columnwidth}{l@{\hspace*{40pt}}l@{\hspace*{40pt}}l}
  \hline
  Label & Effective range & Number of label\\
  \hline
  $T\rm{_{eff}}$ & 3500 K $\sim$ 5500 K & 75,959\\[2pt]
  log \textit{g} & 0.00 dex $\sim$ 4.00 dex & 75,959\\[2pt]
	[Fe/H] & -2.50 dex $\sim$ 0.50 dex & 75,959\\[2pt]
	[M/H] & -2.50 dex $\sim$ 0.50 dex & 75,959\\[2pt]
	[$\alpha$/M] & -0.15 dex $\sim$ 0.40 dex & 75,070\\[2pt]
	[C/Fe] & -0.70 dex $\sim$ 0.40 dex & 75,424\\[2pt]
	[N/Fe] & -0.40 dex $\sim$ 0.70 dex & 74,556\\[2pt]
	[O/Fe] & -0.20 dex $\sim$ 0.70 dex & 75,564\\[2pt]
	[Mg/Fe] & -0.20 dex $\sim$ 0.50 dex & 75,848\\[2pt]
	[Al/Fe] & -0.50 dex $\sim$ 0.50 dex & 74,674\\[2pt]
	[Si/Fe] & -0.20 dex $\sim$ 0.40 dex & 75,730\\[2pt]
	[Ca/Fe] & -0.20 dex $\sim$ 0.40 dex & 75,239\\[2pt]
	[Mn/Fe] & -0.50 dex $\sim$ 0.30 dex & 72,573\\[2pt]
	[Ni/Fe] & -0.20 dex $\sim$ 0.20 dex & 74,575\\[2pt]
  \hline
 \end{tabular*}
\end{table}

It is worth mentioning that when limiting the uncertainty and parameter space, we do not remove a certain star from the dataset just because one of its labels or the uncertainty of a label does not meet the requirements. What we do is to set the label and its uncertainty to NaN, preventing them from affecting the loss function. Other labels of the star that meet the requirements will still participate in the network training.\par
	Although our screening process for each label was relatively independent, when we trained the neural network, all labels as well as their uncertainties were learned simultaneously except for the empty ones. When the network makes predictions for large samples from LAMOST DR8, abundances and uncertainties of abundances of all elements will also be output simultaneously, rather than integrating predictions of individual elemental abundances, so that consistency between the elemental abundances is preserved.\par
\subsubsection{Sample weights}
In order to make the neural network better fit the metal-poor part of the training set, we prepend a weight matrix to the loss function. In this weight matrix, we can edit the weights corresponding to each star. We need to increase the weight of the sparse part of the sample distribution, that is, the metal-poor part. Besides, we should also appropriately reduce the weight of the dense part of the sample distribution, that is, the samples with [Fe/H] between -0.5 dex and 0.1 dex. We divided the [Fe/H] range of the samples in the training set into 30 bins, each with a width of 0.1 dex. Then we count the number of samples in each bin. We found that samples with -1.4 dex < [Fe/H] < -0.9 dex were almost evenly distributed, with an average of about 410 stars in each bin. There are 293 stars in the bin to the left of this interval and 756 stars in the bin to the right of it. There seems to be a gap between the numbers of stars in the two bins and this interval appears to be a buffer, so we use 410 as the weight standard for the metal-poor stars. The weight of samples with [Fe/H] < -1.4 dex is equal to 410 divided by the total number of stars in the bin that the sample is in. As there are only 14 stars with [Fe/H] < -2.4 dex and 38 stars with -2.4 dex < [Fe/H] < -2.3 dex, the weight could become too much for them in this way. To prevent overfitting, we set the maximum weight to be 10. For stars with -0.5 dex < [Fe/H] < 0.1 dex, the weight is set to be 0.95 to reduce their influence on the cost function. For the stars we didn’t mention, the weight defaults to 1.
\begin{figure*}
 \includegraphics[width=\textwidth]{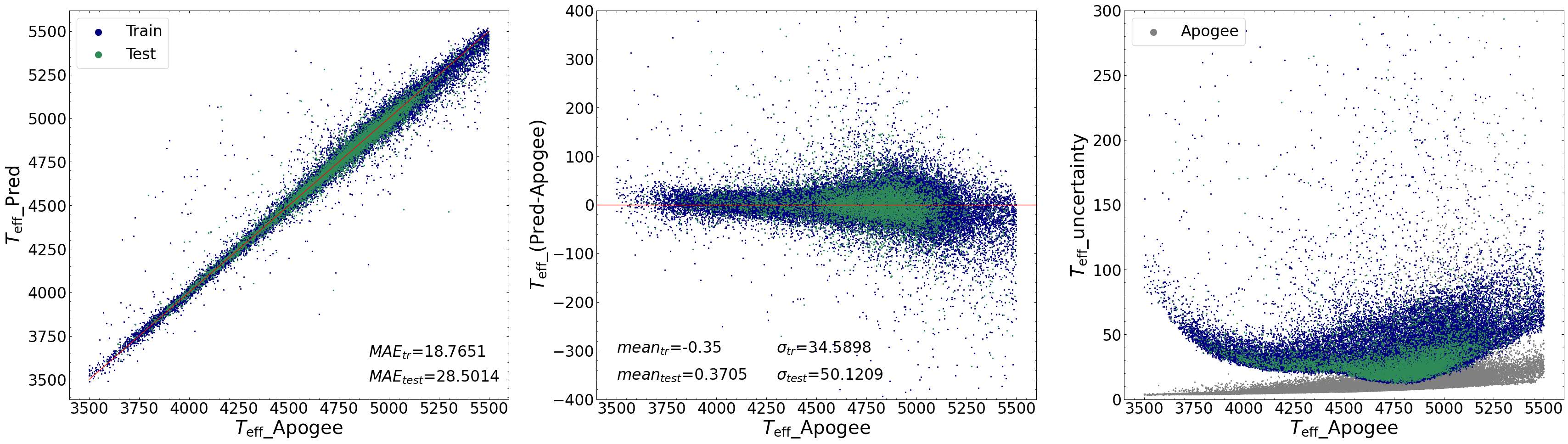}
 \caption{Neural Network performance in predicting effective temperature. The blue and green dots in all the subplots represent the training set and the test set, respectively. MAE in the left subplot is the mean absolute error of the training set and the test set. The arithmetic mean value of the residuals, which represents the bias between neural network predictions and APOGEE labels is listed in the middle subplot along with the standard deviation of the residuals. For the blue and green dots, the right subplot shows the distribution of uncertainties given by the neural network, while the grey dots represent the uncertainty of APOGEE labels for both training and test sets.}
 \label{fig:Teff}
\end{figure*}

\section{Neural Network Performance}
\label{sec:3}
\subsection{Effective temperature and surface gravity}
The predictions of the effective temperature given by the neural network for samples in the training set and test set are shown in Figure~\ref{fig:Teff}. The left subplot is a comparison of the predictions of the neural network (NN) and the APOGEE labels, the x-axis is the effective temperature given by APOGEE and the y-axis is the NN prediction. In this and similar figures that follow, green dots represent stars in the test set while blue dots represent ones from the training set, and it should be noted that the left subplot shows all samples in the test and training sets, while the middle and the right subplots are truncated on the coordinates for a better view. The red line is the diagonal line of the plane, which represents a one-to-one correspondence. In an ideal result, the dots should be distributed around this line. We present the mean absolute error of the training set and the test set in the lower right corner of the subplot. The mean absolute error (MAE) is widely used to measure the accuracy of neural network regression predictions, defined as
\begin{equation}
MAE = \frac{\sum_{i=1}^n\lvert \hat{y}_i-y_i \rvert}{n}
\end{equation}
where $\hat{y}_i$ is the predicted value and ${y}_i$ is the value of the label. The number of samples is denoted as $n$. The mean absolute error is 19 K for the training set and 29 K for the test set, these are very small errors compared to the larger value of the effective temperature. The middle subplot shows the residual between predicted values and APOGEE labels, which is defined as NN predictions minus the values given by APOGEE. The red line is a y=0 baseline where the dots should be distributed around in an ideal result. The mean value presented in the subplot is the arithmetic mean value of the residuals, it represents the systematic bias between neural network predictions and APOGEE labels and it’s only -0.35 K for the training set and 0.37 K for the test set. The standard deviation of the residuals is shown as $\sigma$, it represents the degree of dispersion of the residuals. For effective temperature, the standard deviation of the residuals of training set and test set are 35 K and 50 K respectively. The right subplot shows the distribution of uncertainty of the effective temperatures of the training set, the test set and APOGEE labels which are shown as grey dots, the x-axis is the effective temperatures measured by APOGEE. In the first two figures, there is an obvious phenomenon that the NN predictions for the stars with effective temperature lower than 3700 K tend to be a slightly higher, and for the hotter stars whose effective temperature is higher than 5400 K, the predictions tend to be lower. This is because there are fewer samples at the cold and hot ends but the neural network tends to fit parts with more samples, so the effective temperatures of the colder ones are overestimated and the effective temperatures of the hotter ones are underestimated, we call this phenomenon the boundary effect of neural networks. A similar trend can also be seen in the right subplot, the uncertainties of the predictions become larger at the cold and the hot parts. In general, a prediction with larger deviation from the label tends to have larger uncertainty, we can use the uncertainty to screen out samples with accurate predictions when the neural network is used for large sample predictions and there are not any labels as ground truth. Taken together, we believe that the predictions given by the neural network are reliable between 3700 K and 5400 K, these predicted values can also be further screened using uncertainty as needed.\par
\begin{figure*}
 \includegraphics[width=\textwidth]{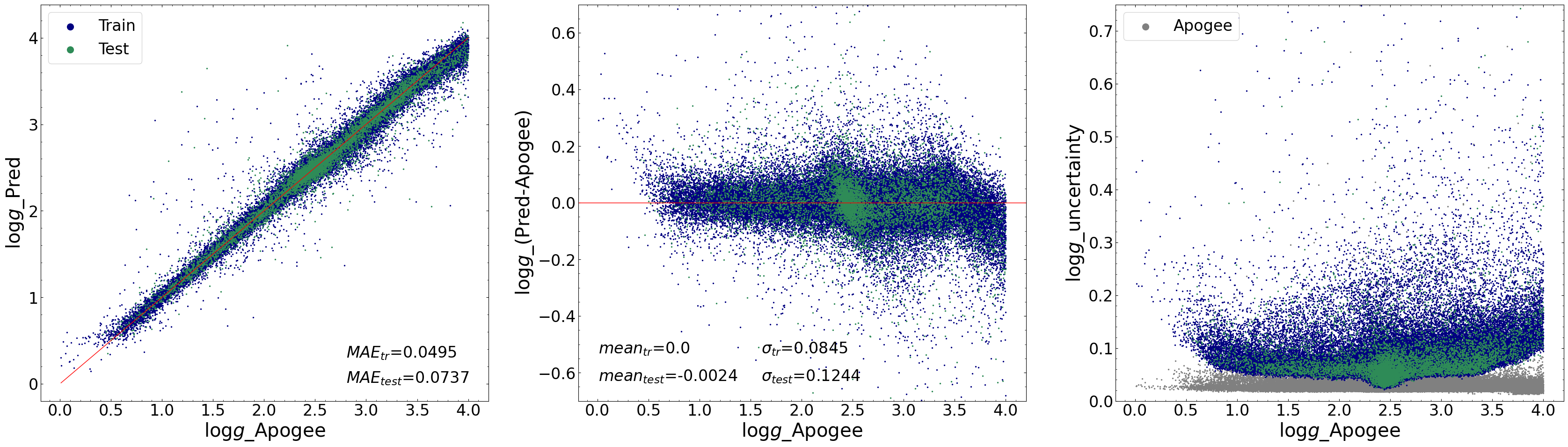}
 \caption{Neural Network performance in predicting surface gravity.}
 \label{fig:logg}
\end{figure*}

Figure~\ref{fig:logg} shows the performance of the neural network in predicting surface gravity. A comparison of the predictions and the labels is shown in the left subplot and the middle one shows the distribution of the residuals. The red diagonal and horizontal lines represent the ideal situations, which are basically impossible to achieve across the entire parameter domain experimentally due to the nonuniformity of the label distribution. The stars with surface gravity lower than 0.6 dex is overestimated and ones with surface gravity higher than 3.5 dex is underestimated. But statistically, the predictions of the neural network have little systematic error with the labels of APOGEE. The mean values of residuals for the training set and the test set are both close zero. The standard deviations of the residuals in the training set and test set are 0.08 dex and 0.12 dex respectively. In the subplot on the right, there is an upward trend in the prediction uncertainty for samples with marginal surface gravity values, which is similar to the case for effective temperature. Although the uncertainty of most predicted values is below 0.2 dex, we recommend using neural network predictions of surface gravity between 0.6 dex and 3.5 dex.

\subsection{Metallicity}
\begin{figure*}
 \includegraphics[width=\textwidth]{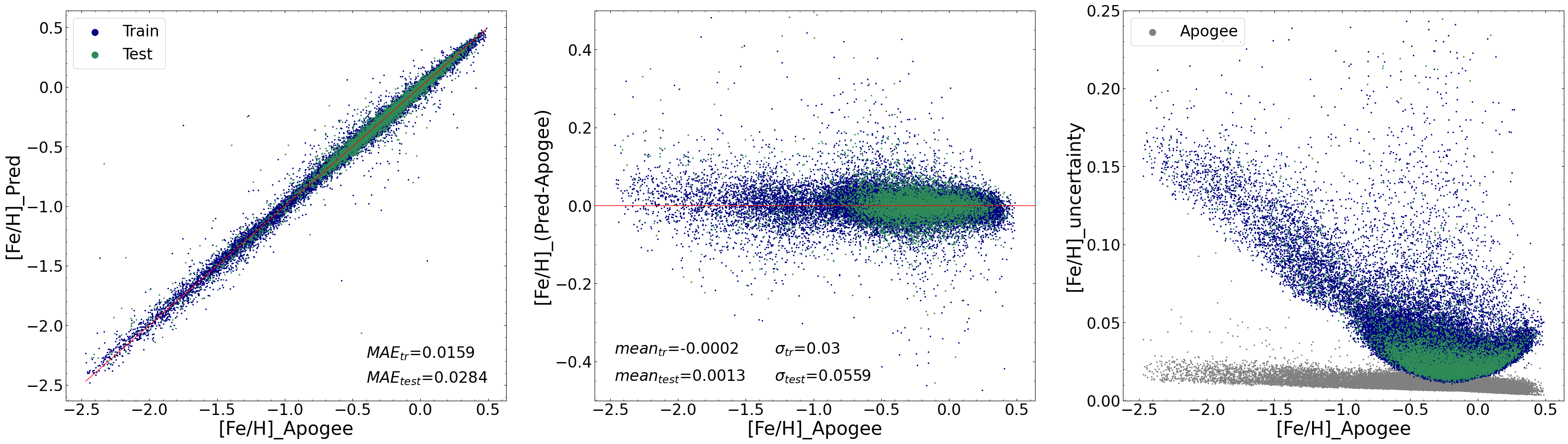}
 \includegraphics[width=\textwidth]{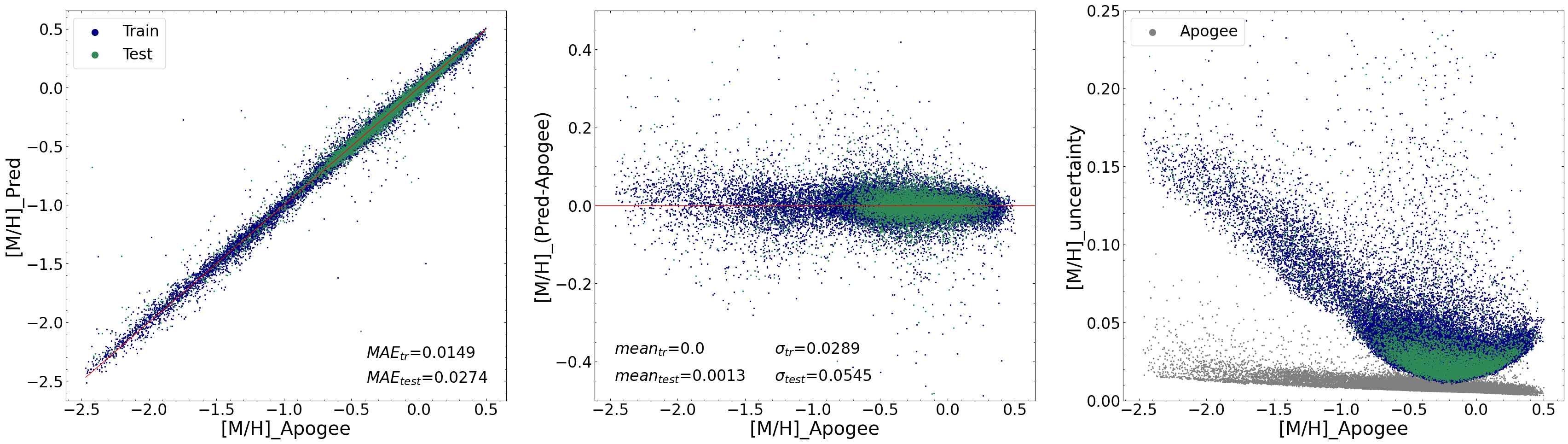}
 \caption{Neural Network performance in predicting [Fe/H] and [M/H].}
 \label{fig:FE_H}
\end{figure*}

There two parameters that characterize metallicity in APOGEE DR17, total metallicity [M/H] and [Fe/H] which is measured directly from Fe lines and is used more commonly. Although [M/H] is generally very close to [Fe/H] because most iron-group elements usually vary in lockstep and iron dominates in metals, we still let the neural network learn both labels at the same time. The first row of Figure~\ref{fig:FE_H} shows the performance of the neural network in predicting [Fe/H] and the second row shows the performance on [M/H]. As we expected, the neural network performed very similarly on the predictions of these two parameters. The difference between the mean absolute errors of neural network predictions for [Fe/H] and [M/H] is negligible both on the training set and the test set. The predictions are accurate for both parameters, since the MAE is 0.02 dex for the training set and 0.03 dex for the test set. In the middle columns of Figure~\ref{fig:FE_H}, there is basically no upturn or downturn even at the metal-poor part. This is because we weight the samples during training, allowing the neural network to better fit the metal-poor part with fewer samples. The systematic difference between the predicted values of [Fe/H] and the APOGEE label is very small, the mean of the residuals is close to zero on both the training and the test set. In the right columns, there is a clear increase in the uncertainty of the prediction as the metallicity decreases. Although samples on the metal-poor part have bigger weights, they are still much less than samples with [Fe/H] between -1.0 dex and 0.5 dex. For [M/H], the situation is similar. Considering that the uncertainty of most of the prediction results is lower than 0.2 dex, we believe that the metal abundance predictions given by the neural network are reliable within the whole sample range. However, the uncertainty can be further limited as needed when using the predicted results.\par
\begin{figure*}
 \includegraphics[width=\textwidth]{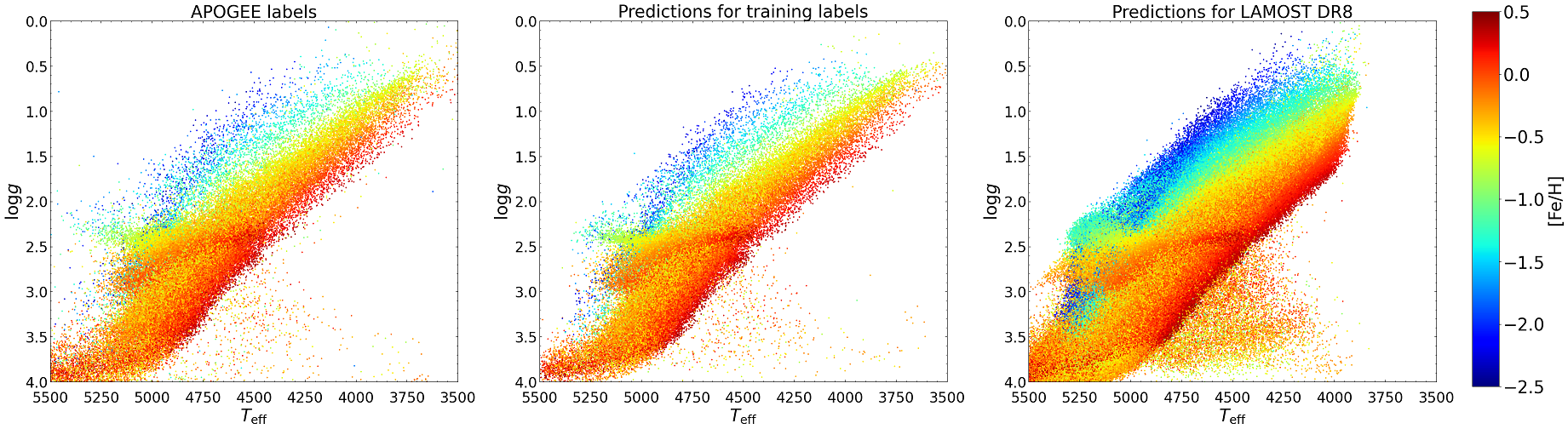}
 \caption{Kiel diagrams for APOGEE labels, Neural Network predictions on the training and the test sets, and Neural Network predictions on target samples. All the subplots are color-scaled by [Fe/H].}
 \label{fig:6}
\end{figure*}

We show the distribution of labels from APOGEE, neural network predictions for training samples and neural network predictions for LAMOST DR8 target samples on the Kiel diagram in Figure~\ref{fig:6}. All the subplots present log \textit{g} vs. $T\rm{_{eff}}$, and are color-scaled by [Fe/H]. Both the subplot in the middle and the subplot on the right are very similar to the subplot on the left, which shows that the predictions of atmospheric parameters given by the neural network are very accurate and have good consistency with APOGEE labels. Note that the red clump (RC) stars are similar prominent populations in all subplots, and metal-poor RC stars have slightly lower log \textit{g} than the relatively metal-rich ones (see different colors with $T\rm{_{eff}}$ = 5000 K ), which was suggested by \citet{zhao2001high} based on high-resolution spectra obtained at Xinglong station \citep{zhao2001coude} for 39 red clump stars. This consistency shows that the parameters estimated by the neural network for LAMOST DR8 samples are reliable.\par
The prediction range of the effective temperature of the neural network can be as low as 3500 K, but there are very few stars with effective temperature below 3800 K in the LAMOST DR8 sample, so there are some blanks in the low temperature region of the right subplot. For all the subplots, we intercept the part where the parameters satisfy 3500 K < $T\rm{_{eff}}$ < 5500 K, 0.0 dex < log \textit{g} < 4.0 dex because these are the parameter ranges defined when we select samples using LAMOST Stellar Parameter Pipeline and it is also the range where our training samples lie. In the actual predictions of the neural network, there are a small number of stars whose effective temperature is predicted to be higher than 5500 K and the surface gravity higher than 4.0 dex. The prediction results for this part of the stars can be regarded as the extrapolation of the neural network, which we do not recommend using. Most of the predictions that lie in the parameter space of the training set are fairly reliable.

\subsection{$\alpha$-elemental abundances}
\label{sec:3.3}
\begin{figure*}
 \includegraphics[width=\textwidth]{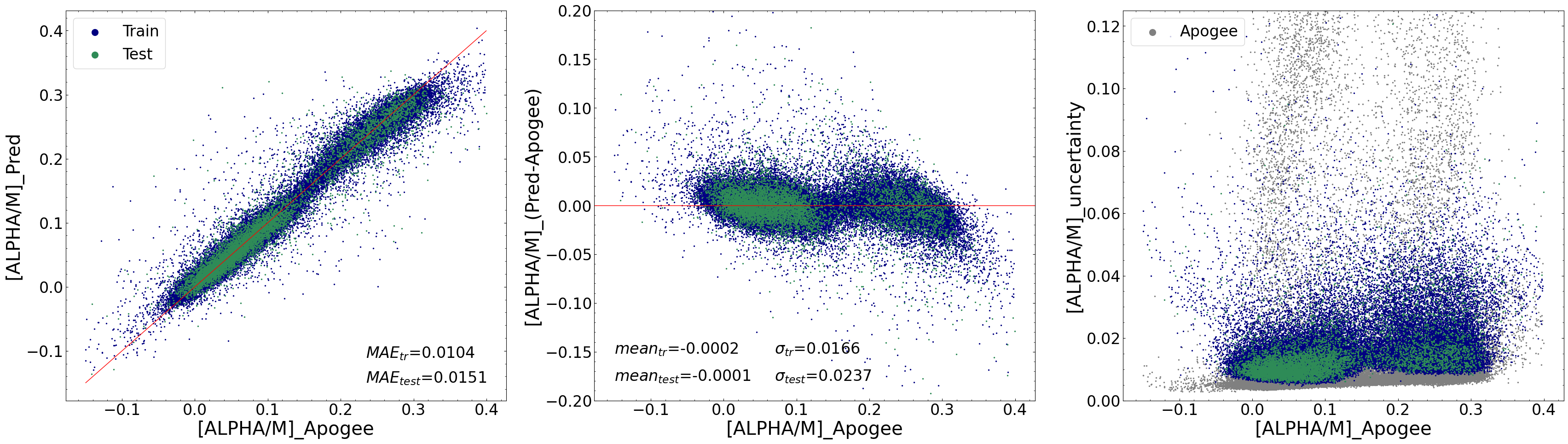}
 \caption{Neural Network performance in predicting [$\alpha$/M].}
 \label{fig:7}
\end{figure*}

For $\alpha$ elements, APOGEE DR17 catalog provides the total $\alpha$ abundance [$\alpha$/M] along with six individual $\alpha$ elements O, Mg, S, Si, Ca, Ti. In addition to [$\alpha$/M], we selected O, Mg, Si, Ca from the six $\alpha$ elements given by APOGEE, which are more accurate in measurement and have better training results, for the neural network to learn. Figure~\ref{fig:7} shows the performance of our neural network in predicting the total abundance of $\alpha$ elements. The biggest difference between Figure~\ref{fig:7} and previous similar plots is that there are two distinct clusters of predicted values. This is because the $\alpha$ abundance distribution of the sample has a bimodal structure, one is a low $\alpha$ abundance part dominated by thin disk stars, and the other is a high $\alpha$ abundance part dominated by thick disk stars. Since the neural network is more inclined to fit the densely distributed part of the samples, the predictions are affected by this bimodal structure. For thin disk, stars with very low $\alpha$ abundance are overestimated, the relatively $\alpha$ rich portion are underestimated but not as much as the low $\alpha$ abundance part. For thick disk, the situation is reversed. This is also the case for subsequent predictions of individual $\alpha$ elements. It seems unavoidable, since we cannot have all elements distributed relatively uniformly at the same time, we only weighted the element iron, which we consider to be the most important. Nonetheless, our results are statistically good. The MAE of the training set is 0.01 dex and it is 0.02 dex for the test set. This means that on average, each prediction deviates very little from the label. The mean value of the residuals is nearly zero for both training and test sets, which means that there is basically no systematic error between the predictions and the APOGEE labels. In the subplot on the right, we can see some labels from APOGEE with great uncertainty, but they do not affect the training results much, because the effect of these labels will be weakened by the loss function. We recommend using predicted values between -0.03 dex and 0.32 dex, predictions outside this range with less uncertainty can be used as reference. It is worth mentioning that, starting with [$\alpha$/M], there will be some stars with empty labels in the data set, they will not appear in the plots and won’t participate in the calculation of the MAE or the residuals. The numbers of stars shown in similar figures may be different, and the specific numbers are shown in Table~\ref{tab:1}.\par
\begin{figure*}
 \includegraphics[width=\textwidth]{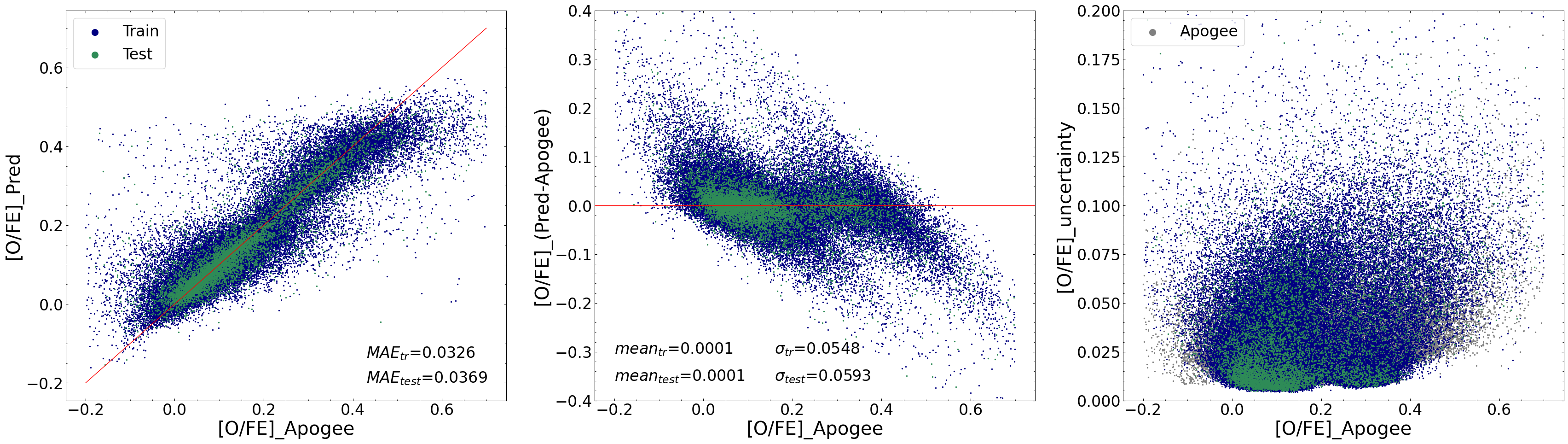}
 \includegraphics[width=\textwidth]{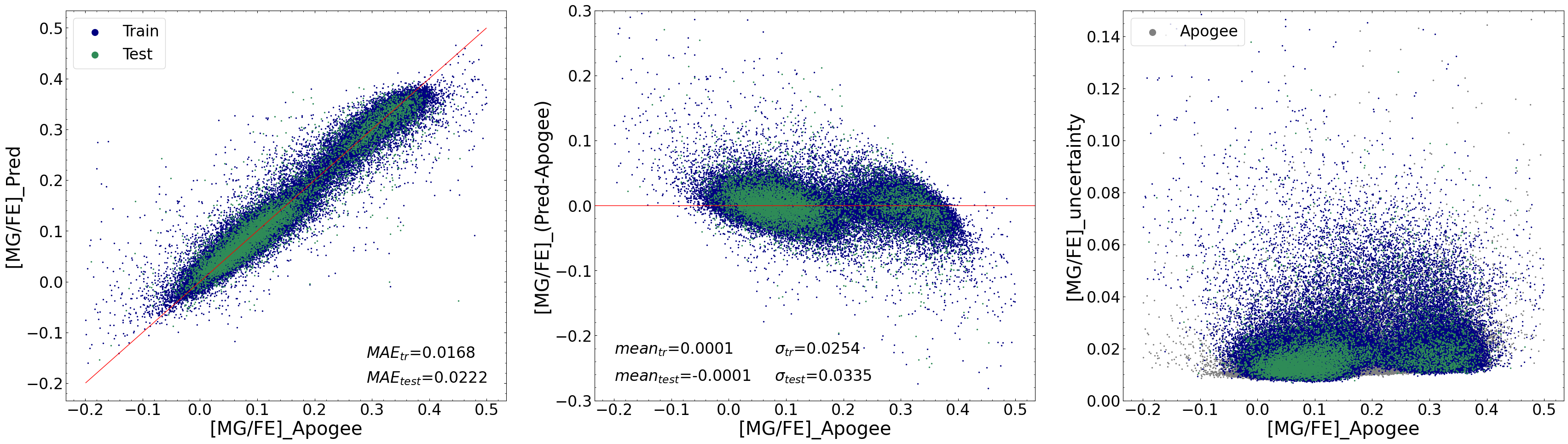}
 \includegraphics[width=\textwidth]{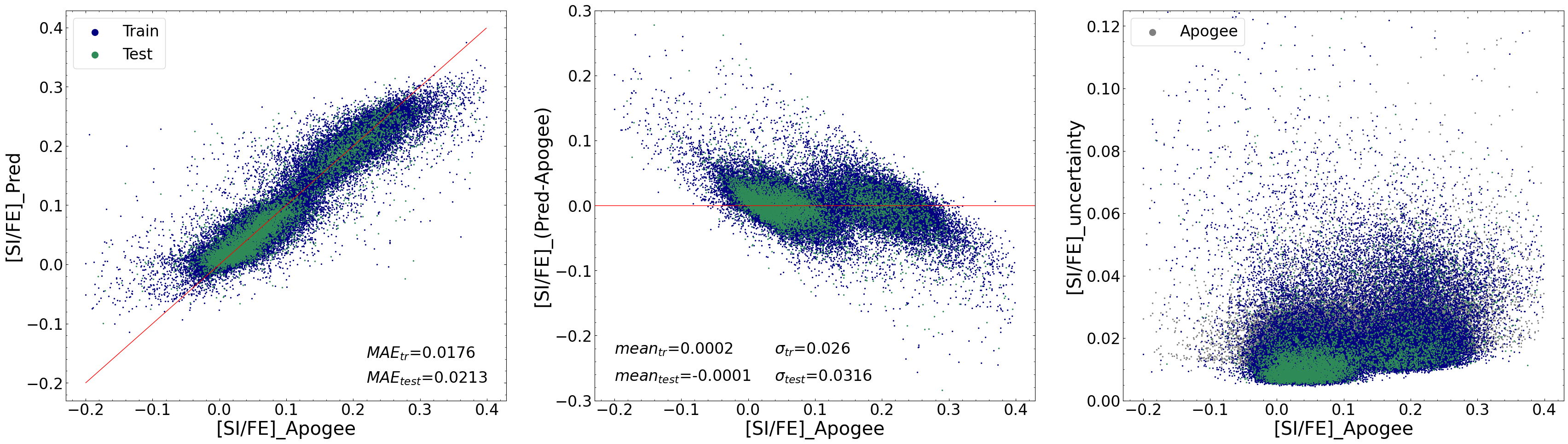}
 \includegraphics[width=\textwidth]{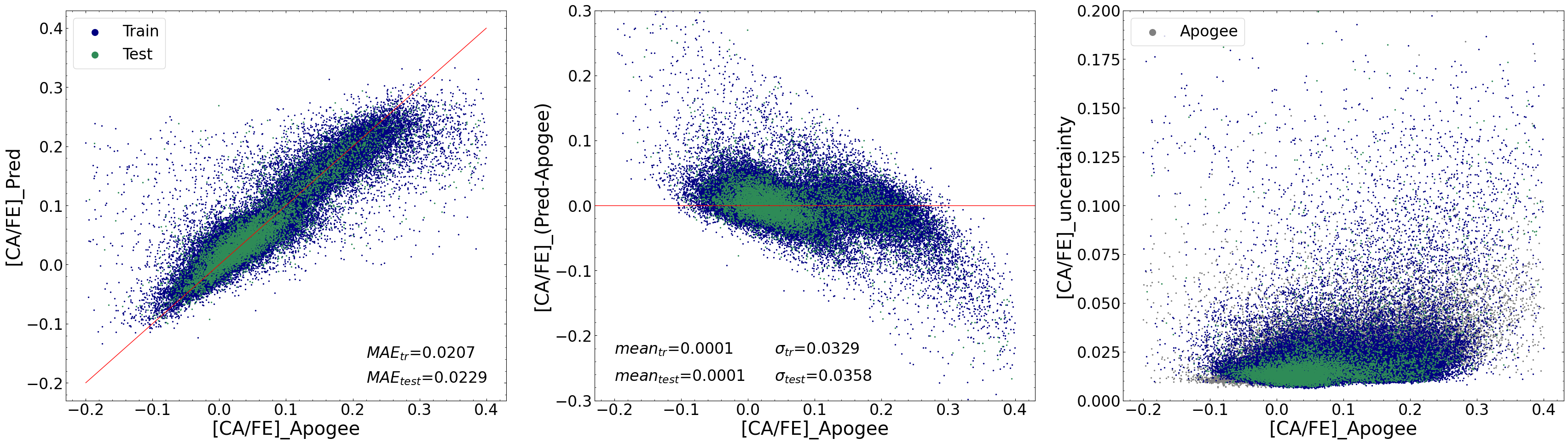}
 \caption{Neural Network performance in predicting [O/Fe], [Mg/Fe], [Si/Fe] and [Ca/Fe].}
 \label{fig:8}
\end{figure*}

Each row from top to bottom in Figure~\ref{fig:8} shows the performance of the neural network in predicting oxygen, magnesium, silicon and calcium abundances, respectively. The neural network's performance was generally similar to that in predicting total $\alpha$ abundance, but the quality of the predictions varies for different elements. The neural network's prediction of magnesium abundance is the most accurate among the four $\alpha$ elements, followed by silicon and calcium. The prediction results of oxygen abundance by the neural network are relatively poor. In the left subplot for oxygen, we can see two traces that are diffuse and different from most of the samples. The correlation between the predicted values given by the neural network and the APOGEE labels of these samples is weaker than that of most of the samples, but they are not completely uncorrelated. In the middle subplot, this appears as two slanted traces. This is the result of a combination of what we call the boundary effect of neural networks and the bimodal structure of the sample distribution, which can also be seen in the predictions for silicon and calcium, but not as severe as oxygen. The prediction results for magnesium did not appear to be affected much, the neural network performed very similarly in predicting the abundance of magnesium as it did in predicting [$\alpha$/M]. In our low-resolution spectra from LAMOST DR8, the Mg\,{\sevensize I} b triplet are basically distinguishable, but Ca lines and Si lines are mixed with other lines. As for oxygen, it is difficult to obtain its abundance even for high-resolution spectra. Among the four $\alpha$ elements learned by the neural network, magnesium has the most obvious features, which makes the prediction of magnesium the most accurate and the prediction for the abundance of total alpha elements will be dominated by magnesium. This also explains why our neural network performs most similarly in predicting magnesium abundance as it predicted [$\alpha$/M]. From the perspective of MAE, the mean of residuals and the standard deviation of residuals, we find that the neural network performs statistically good, since these values are very small. There are little systematic errors between the NN predictions and the APOGEE labels for these four elements.\par
\begin{figure*}
 \includegraphics[width=\textwidth]{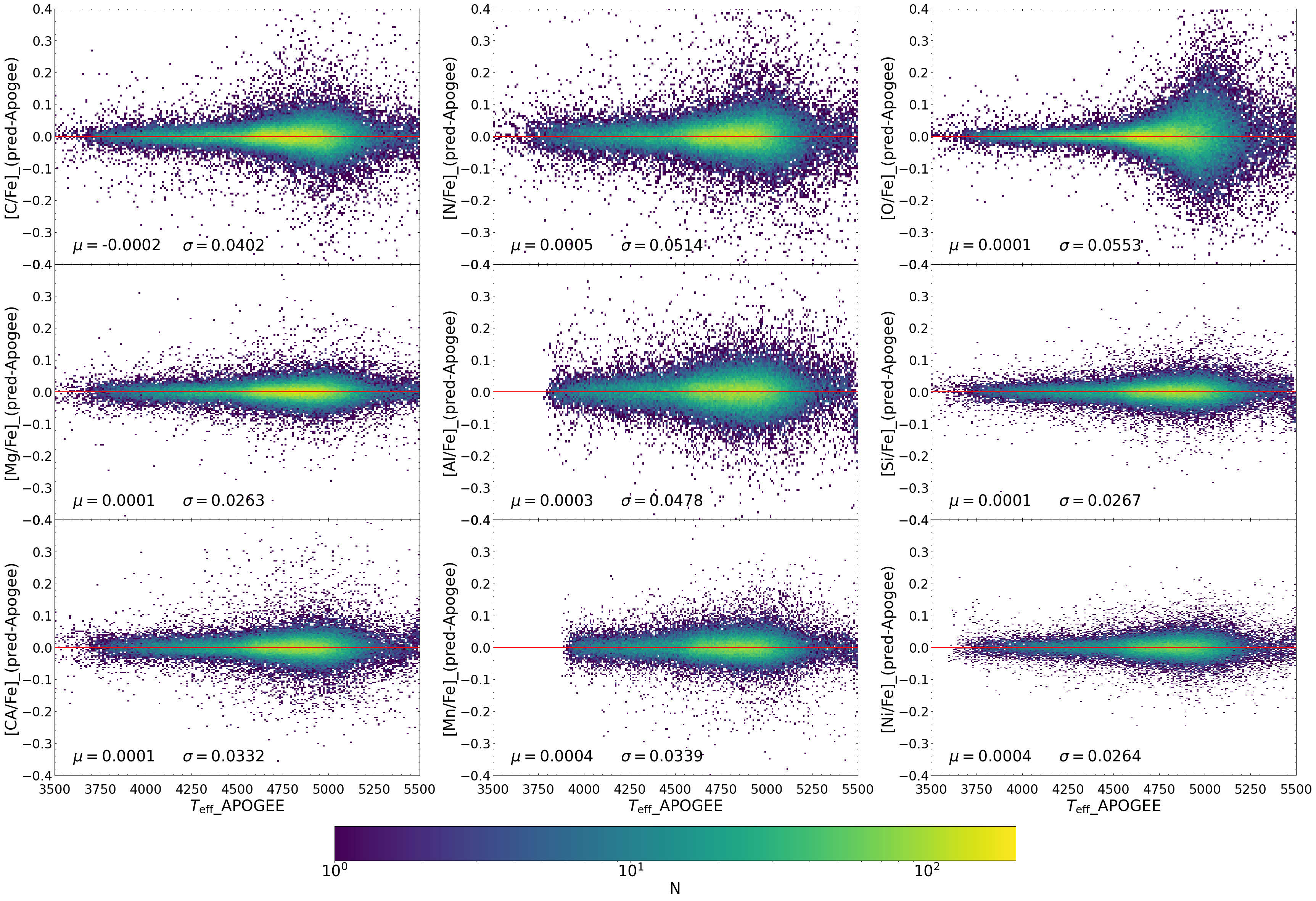}
 \caption{The distribution of the residuals between neural network predictions and APOGEE labels on the effective temperature. The x-axis is the effective temperature measured by APOGEE, the y-axis is the residual which is defined as neural network predictions minus APOGEE labels. The color represents the number of stars in each bin.}
 \label{fig:9}
\end{figure*}

\begin{figure*}
 \includegraphics[width=\textwidth]{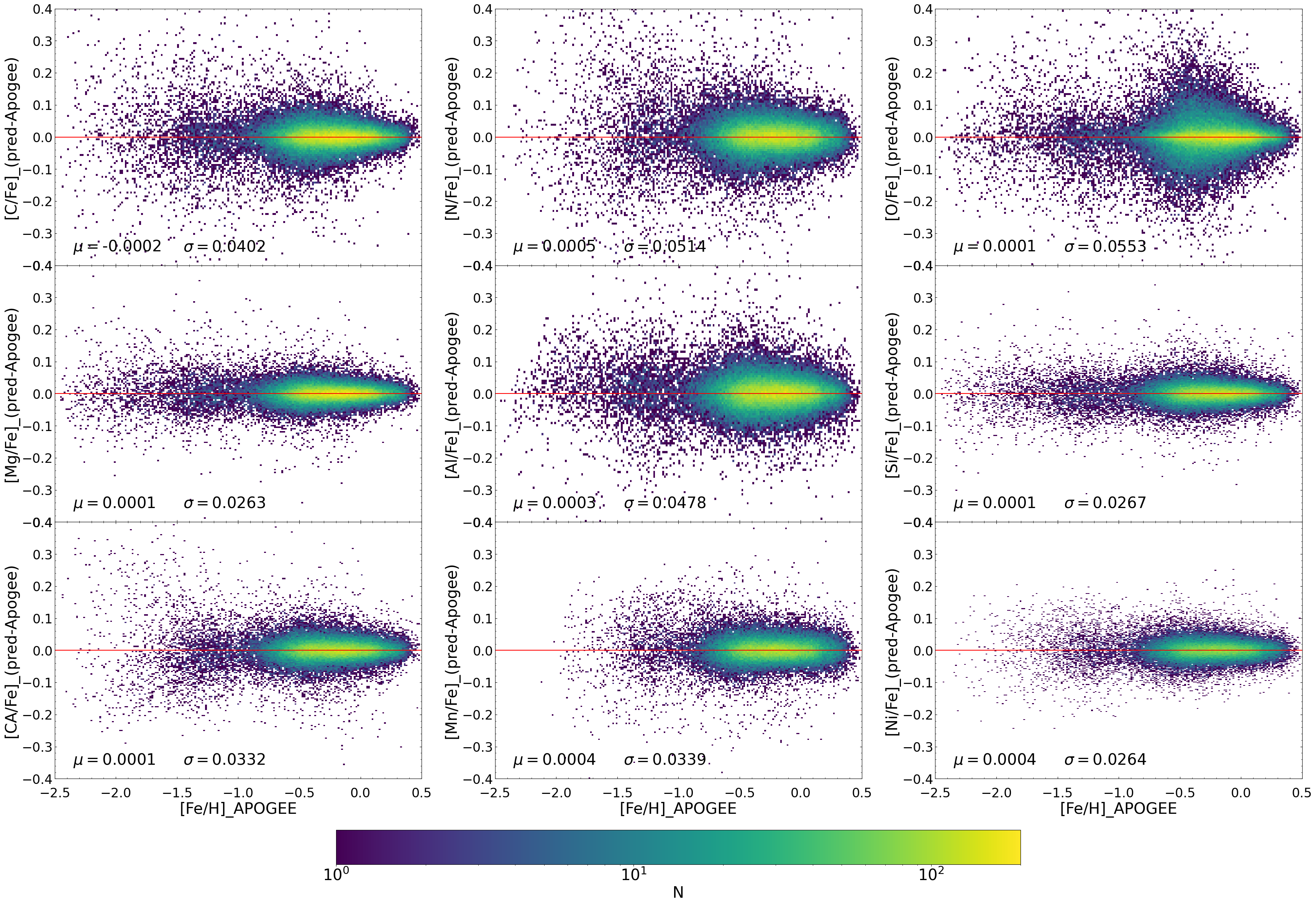}
 \caption{The distribution of the residuals between neural network predictions and APOGEE labels on the iron abundance. The x-axis is [Fe/H] measured by APOGEE, the y-axis is the residual which is defined as neural network predictions minus APOGEE labels. The color represents the number of stars in each bin.}
 \label{fig:10}
\end{figure*}

We show the distribution of residuals versus the effective temperature in Figure~\ref{fig:9}. In each subplot, the x-axis is the effective temperature given by APOGEE, the y-axis is the residuals between neural network predictions and APOGEE labels, the color represents the number of samples in each bin. In the lower left corner of each subplot, we show the mean and standard deviation of the residuals for each element, labelled as $\mu$ and $\sigma$ respectively. We can visualize how the accuracy of the neural network prediction varies with effective temperature in Figure ~\ref{fig:9}, and we can also understand the effective temperature distribution of samples with different non-empty labels. Figure~\ref{fig:10} is similar to Figure ~\ref{fig:9}, the only difference is that the x-axis of Figure~\ref{fig:10} represents iron abundance, it shows the distribution of residuals versus [Fe/H].\par
In Figure~\ref{fig:9}, the residuals for the four $\alpha$ elements are all around zero over the entire effective temperature range from 3500 K to 5500 K. The scatter of the residuals of oxygen is very small when the effective temperature is below 4500 K, but it increases rapidly when the effective temperature is higher than 4500 K. This is because the uncertainties of the labels from APOGEE rise significantly after the effective temperature reach 4500 K, where the uncertainties have been fairly low until then. The rise of temperature will lead to the decomposition of OH, which makes oxygen abundances difficult to be measured, since they are derived from OH lines in the APOGEE survey \citep{jonsson2020apogee}. This part of the stars also contribute to the similar increase in the scatter of the residuals of oxygen in Figure~\ref{fig:10}. As for the other three elements, the scattering changes are relatively gentle, and the standard deviations of their residuals are small. In Figure~\ref{fig:10}, it is obvious that when [Fe/H] is below -1.5 dex, the number of samples with non-empty [Ca/Fe] label drops sharply, which causes the scatter of the residuals to rise in the metal-poor part. For this reason, we consider the calcium abundances of metal-poor stars given by the neural network to be less reliable. They can be used as reference but should be treated carefully. But statistically speaking, the neural network did a decent job of predicting calcium abundance, as the mean of the residuals is fairly small and the standard deviation is 0.03 dex. For the other three elements, the reduction in the number of samples of metal-poor samples is also present but not as severe as for calcium. Their residuals seem to be all around zero over the [Fe/H] range from -2.5 dex to 0.5 dex. In summary, the neural network's predictions for these four alpha elements are fairly accurate. The recommended ranges for the elements are list in Table~\ref{tab:2}. These ranges are subjectively selected, the predictions given by the neural network can be further screened using uncertainty given by the neural network.
\begin{table}
 \caption{Recommended ranges for elemental abundances predicted by the neural network. These ranges are selected subjectively and they are for reference only. It is best to use the prediction uncertainties given by the neural network to select reliable predictions.}
 \label{tab:2}
 \begin{tabular*}{\columnwidth}{l@{\hspace*{40pt}}l@{\hspace*{40pt}}l}
  \hline
  Elements & Lower limit & Upper limit\\
  \hline
  Oxygen & -0.05 dex & 0.45 dex\\[2pt]
  Magnesium & -0.05 dex & 0.38 dex\\[2pt]
	Silicon & -0.05 dex & 0.32 dex\\[2pt]
	Calcium & -0.08 dex & 0.28 dex\\[2pt]
	Carbon & -0.60 dex & 0.22 dex\\[2pt]
	Nitrogen & -0.20 dex & 0.50 dex\\[2pt]
	Aluminum & -0.40 dex & 0.32 dex\\[2pt]
	Manganese & -0.26 dex & 0.20 dex\\[2pt]
	Nickel & -0.05 dex & 0.12 dex\\[2pt]
  \hline
 \end{tabular*}
\end{table}

\subsection{Light elements, odd Z elements and iron-peak elements}
In addition to the elements described above, the neural network also gives the abundances of carbon, nitrogen, aluminum, manganese, and nickel, simultaneously. Carbon, nitrogen and oxygen are very important elements in the process of stellar evolution, and we have already discussed oxygen as an $\alpha$ element in section~\ref{sec:3.3}. Aluminum is an odd Z element, aluminum abundance is often used to study the chemical patterns of stellar targets, as are the abundances of the iron-peak elements manganese and nickel (e.g., \citealt{das2020ages,feuillet2021selecting}). Unfortunately, the spectral lines of these elements are indistinguishable in most of the low-resolution spectra. For this reason, what our neural network has learnt doesn’t seems to be one-to-one features, but some kind of correlation through which the abundances of the elements can be inferred by the model. But for statistical results, this is enough.\par
\begin{figure*}
 \includegraphics[width=0.9\textwidth]{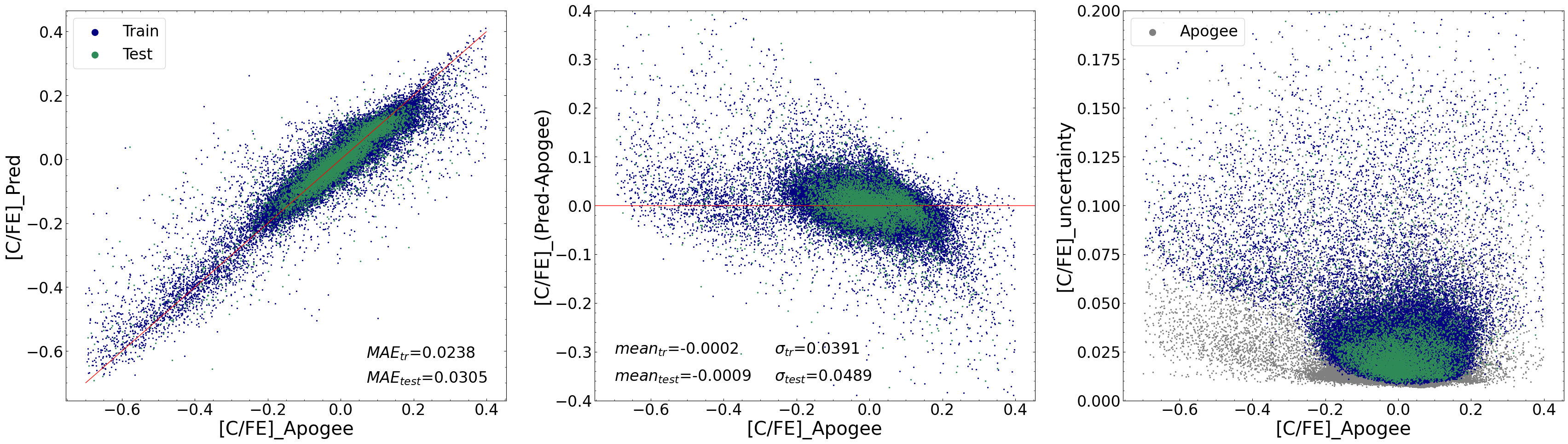}
 \includegraphics[width=0.9\textwidth]{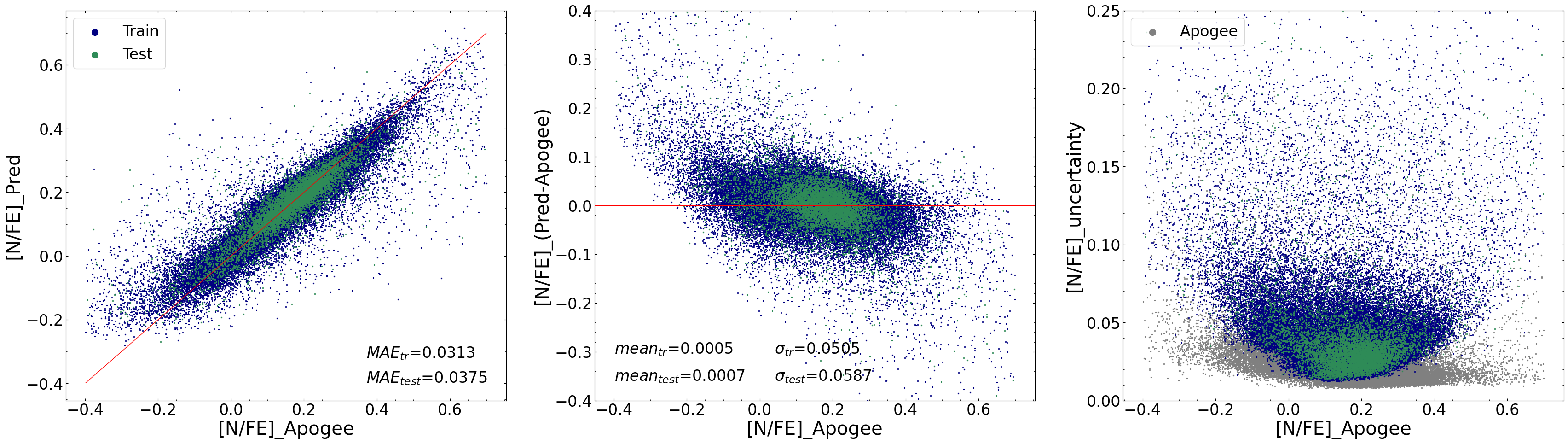}
 \includegraphics[width=0.9\textwidth]{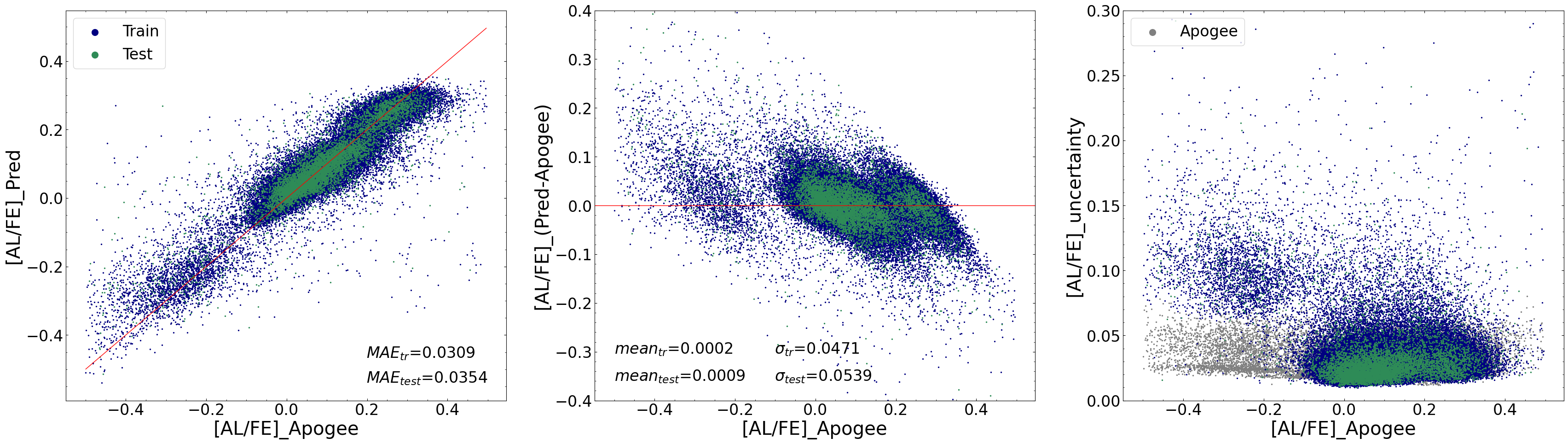}
 \includegraphics[width=0.9\textwidth]{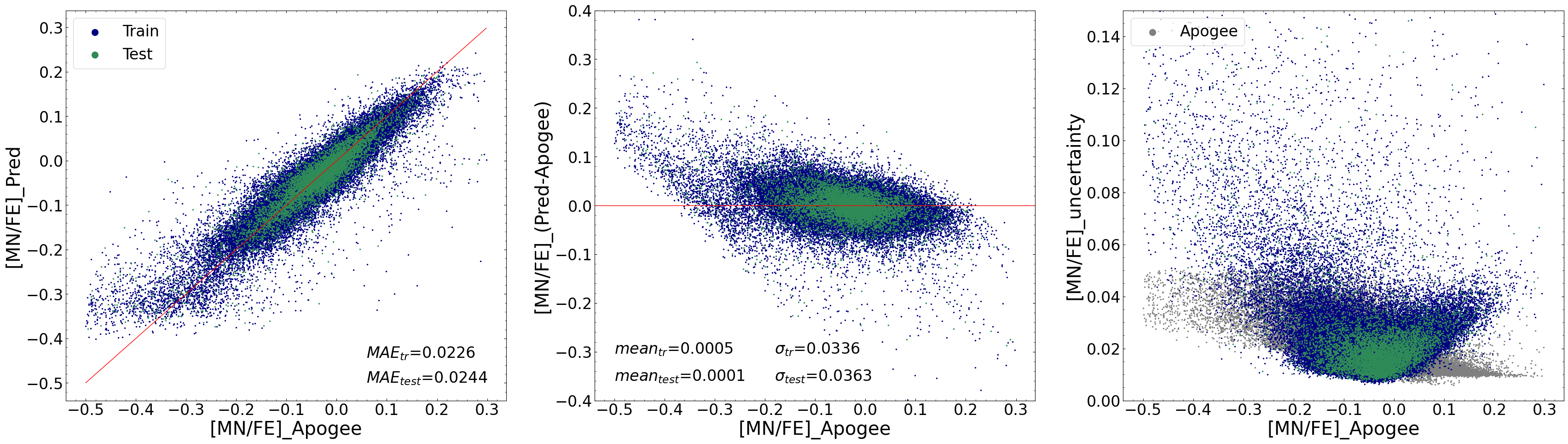}
 \includegraphics[width=0.9\textwidth]{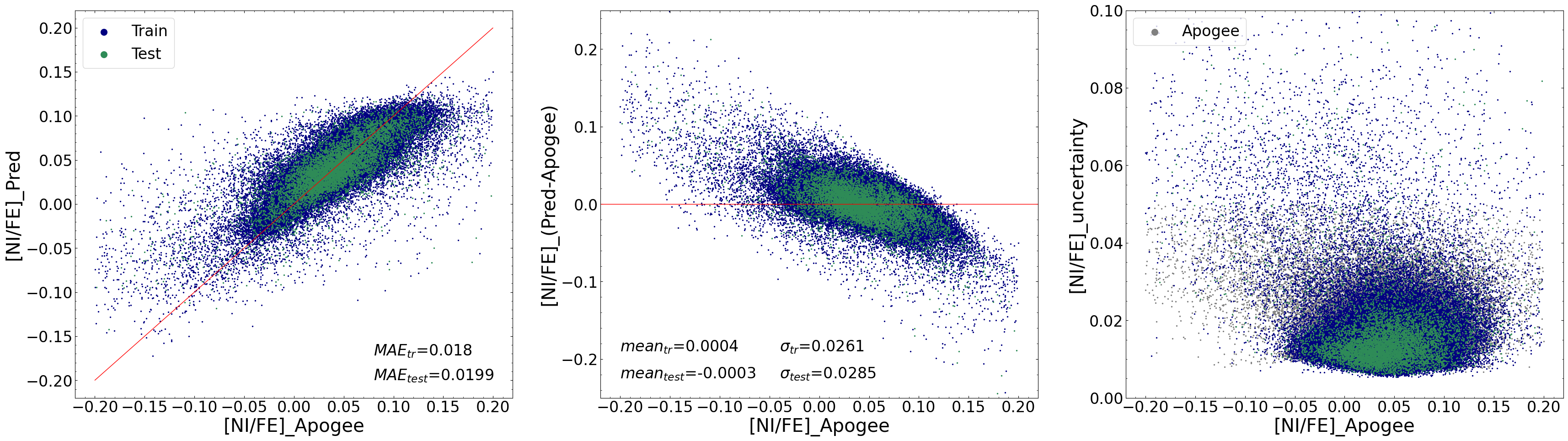}
 \caption{Neural Network performance in predicting [C/Fe], [N/Fe], [Al/Fe], [Mn/Fe] and [Ni/Fe].}
 \label{fig:11}
\end{figure*}

From top to bottom, Figure~\ref{fig:11} shows the performance of the neural network in predicting [C/Fe], [N/Fe], [Al/Fe], [Mn/Fe] and [Ni/Fe]. The neural network's predictions performed statistically well. Whether on the training set or the test set, the mean absolute error of the neural network prediction for all the abundances is less than 0.04 dex. There is no systematic error between the predicted and reference values of the abundances of these five elements, since the mean values of the residuals are all close to zero. Inevitably, there are some stars whose abundances are overestimated or underestimated, which is usually caused by the boundary effect of the neural network and the lack of distinguishable lines in our spectra. But generally, the stars who have large residual always come with a relatively large uncertainty, which can be used to get rid of the less accurate predictions.\par
As the effective temperature decreases, the measurement of atmospheric parameters and element abundances becomes difficult. In Figure~\ref{fig:9}, there is a significant break at about 3900 K for manganese, similar breaks can be seen for aluminum at 3800 K and nickel at 3600 K. For these three elements, there is no sample with non-empty label when the effective temperature is lower than a threshold. For the other two, the number of samples with non-empty label decreases as the effective temperature become lower, but without significant break. Different elements respond differently to temperature, and some elements may not have available spectral lines within a certain temperature range. Even though, in the valid temperature range covered by the samples with non-empty labels, the predicted values are relatively accurate. The mean values of all the residuals are very small and there is no sign of over or underestimation of any abundance. For stars with very low effective temperature in LAMOST DR8 catalog, the neural network will still predict the abundance of all the elements simultaneously, but the predictions for aluminum, manganese and nickel could be extrapolations which we don’t recommend. Since there are only few stars with effective temperature lower than 3800 K, these extrapolations won’t have much impact on the accuracy of our predictions. \par
Things become different as we can see in Figure~\ref{fig:10}. The lack of samples with non-empty label become severe when it comes to the metal-poor end, and the scatter of the residuals become larger which is not the case as we discussed in Figure~\ref{fig:9}. The increase of the residuals scatter is particularly significant in nitrogen and nickel. Manganese is in severe lack of samples when the iron abundance is lower than -1.5 dex. As the metallicity decreases, all but the hydrogen and helium lines in the spectrum become weaker, so the measurement of the abundances of metal-poor stars is difficult even for high-resolution spectra. The uncertainty of our training label from APOGEE DR17 is relatively higher when the metal abundance of the sample is low. On the other hand, the weakening of the spectral lines of the elements makes it a harder job for neural network to extract features from low-resolution spectra. These difficulties are foreseeable, but hard to avoid. Allowing the neural network to train with missing values has already largely improved the sample size of metal-poor stars for several elements, but do not have much help on elements like manganese and nickel. To solve these problems, we can resort to the uncertainties of the predictions given by the neural network. The predictions of the abundances of metal-poor stars generally come with large uncertainties compare to the stars with [Fe/H] > -1.5 dex, we can set a threshold on these uncertainties as needed to remove stars whose predictions are potentially less accurate. In addition, we also present recommended ranges for the elements discussed above in Table~\ref{tab:2}, the ranges are subjectively selected for reference.
\subsection{Prediction uncertainties}
In practical applications, the accuracy of neural network predictions will be affected by many factors, a good neural network model should be robust enough to deal with these challenges. For the model used in this study, the prediction uncertainty provides a realistic estimate of the precision \citep{leung2019deep}. We may choose to use predictions with lower uncertainties for more reliable results.\par
\begin{figure}
 \includegraphics[width=\columnwidth]{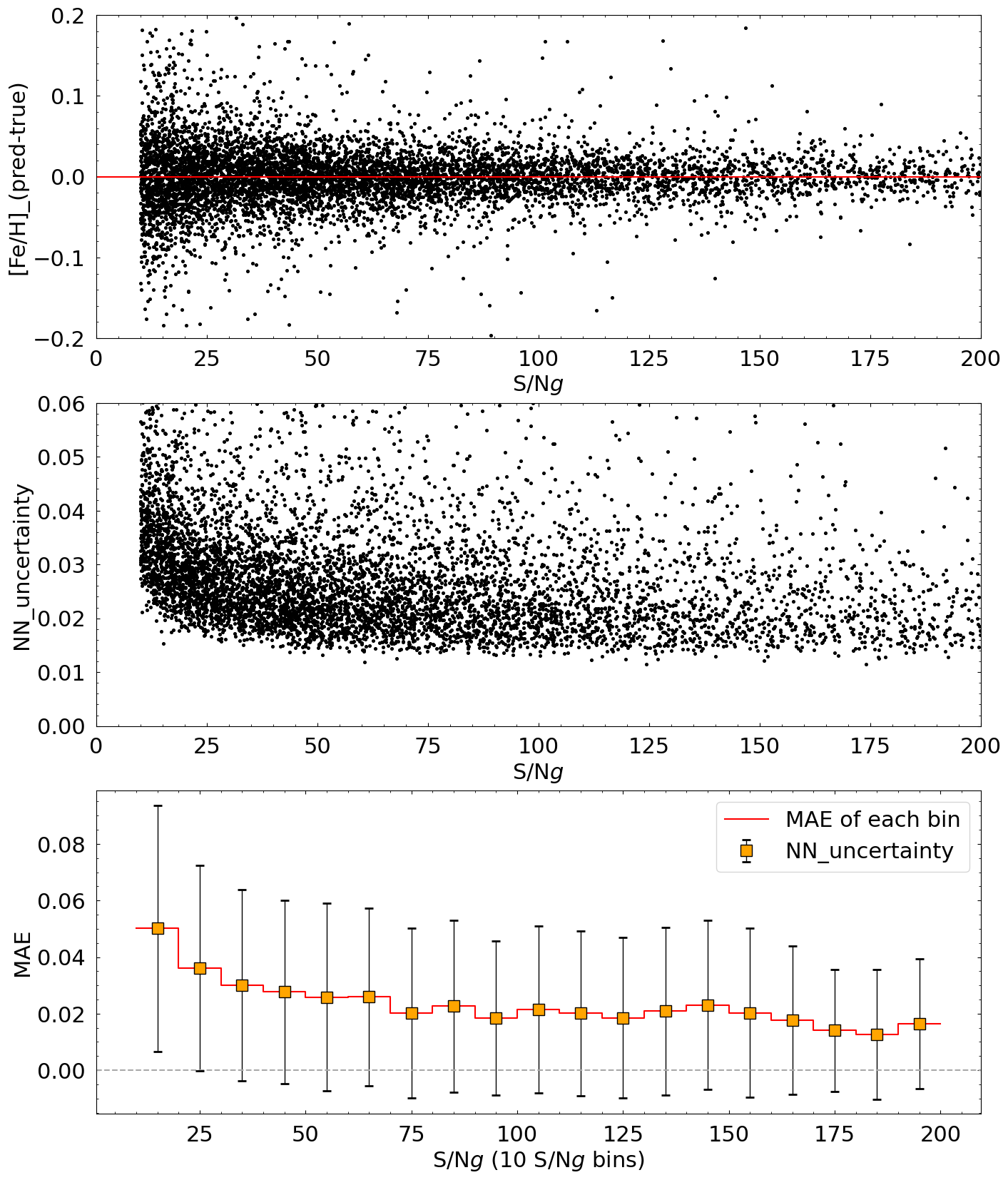}
 \caption{The effect of signal-to-noise ratio on prediction accuracy. The upper and the middle subplots show the distribution of the residuals and the prediction uncertainties as a function of S/N\textit{g}, respectively. The lower subplot shows the mean absolute error as a function of S/N\textit{g}, the error bar represents the mean prediction uncertainty in each bin.}
 \label{fig:Uncertainty1}
\end{figure}

The signal-to-noise ratio has a certain impact on the prediction results, we take [Fe/H] as an example in Figure~\ref{fig:Uncertainty1}. We analysed the samples with S/N\textit{g} of 10 to 200 in our test set. The upper subplot shows the residuals between predictions and reference values distributed with S/N\textit{g}. Obviously, as the signal-to-noise ratio increases, the difference between the overall predicted value and the reference value becomes smaller, since the influence of random noise on spectral features is weakened. The middle subplot shows the distribution of the uncertainties given by the neural network as a function of S/N\textit{g}, the uncertainties also show a tendency to decrease when S/N\textit{g} becomes larger. We further studied the samples in the lower subplot with a signal-to-noise ratio of 10 as a bin. The mean absolute error of each bin is plotted by the red line and also at the centre of the section, the error bar represents the mean total uncertainty of the predictions in each section. The mean absolute error becomes smaller when S/N\textit{g} increases and the mean total uncertainty of the predictions shows a decreasing trend, which has been already seen in the middle subplot. Inevitably, random noise will affect the stability of the model to a certain extent. But in general, the stability of the neural network model is not bad and it will perform better in the case of high signal-to-noise ratio.\par
\begin{figure}
 \includegraphics[width=\columnwidth]{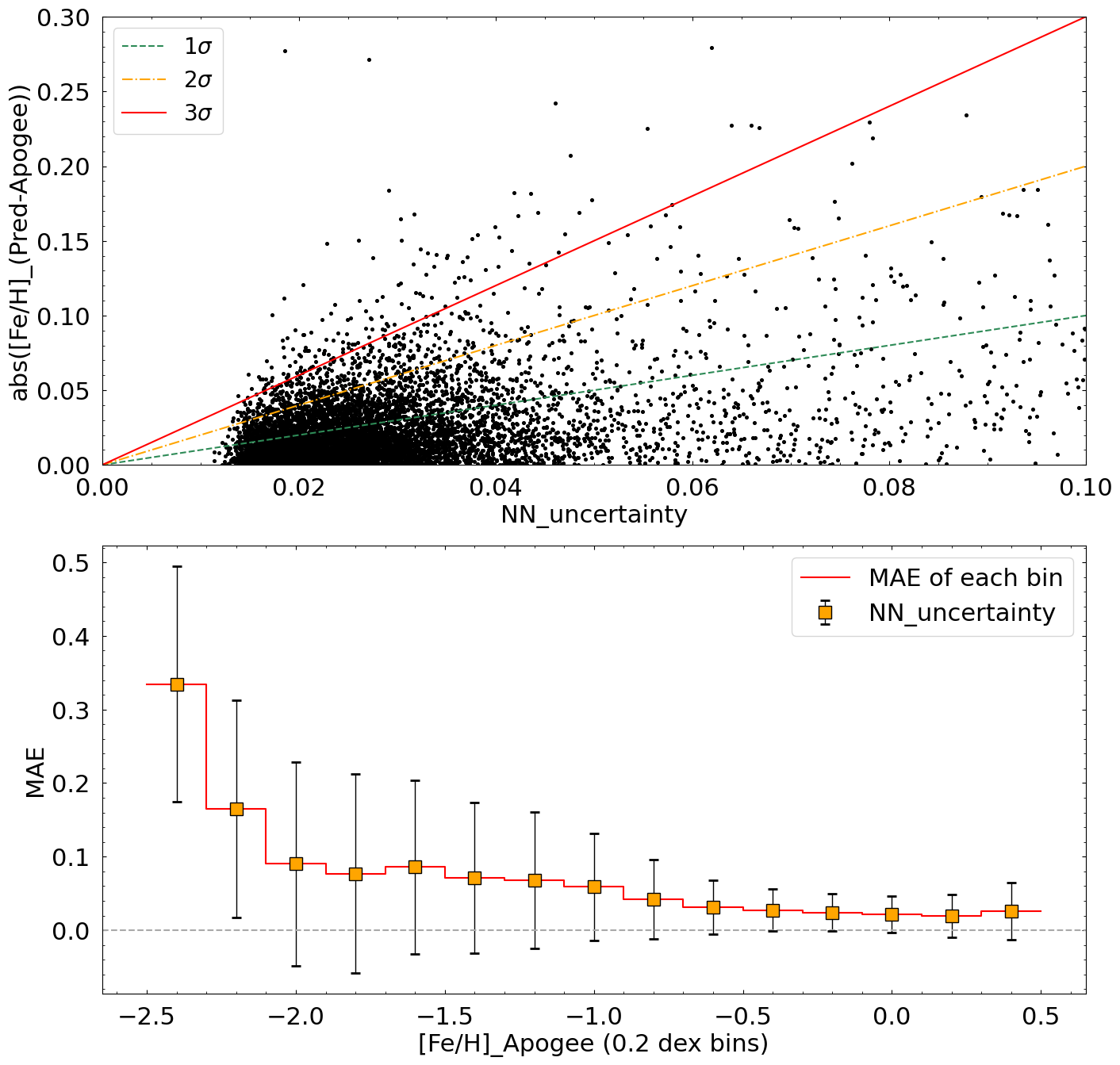}
 \caption{The relation between the uncertainty given by the neural network and the residual. The upper subplot shows the distribution of the absolute value of the residuals between NN predictions and APOGEE labels as a function of prediction uncertainty. The lower subplot shows the MAE as a function of the reference value of [Fe/H], the error bar represents the average value of the uncertainties in each bin.}
 \label{fig:Uncertainty2}
\end{figure}

The distribution of training samples is also an important factor affecting the prediction results. Still taking [Fe/H] as a typical example, Figure~\ref{fig:Uncertainty2} shows the relation between the prediction uncertainty and the residual for samples in the test set. The x-axis of the upper subplot is the uncertainty given by the neural network and the y-axis is the absolute value of the residual between NN prediction and APOGEE label. The different colored lines represent deviations up to 1, 2 and 3 $\sigma$, respectively. The residuals and the uncertainties are positively correlated, and the vast majority of the predictions are within 3 $\sigma$ deviations from the APOGEE labels, which shows that the prediction uncertainty is self-consistent. A similar correlation can also be found between the residuals and APOGEE uncertainties, suggesting that inaccurate labels can affect the performance of the neural network. In the lower subplot, we show the MAE as a function of the reference value of [Fe/H] and mean prediction uncertainties as error bars. With the increase of [Fe/H], there is a very significant decrease in MAE, especially before -0.5 dex. The prediction uncertainty also shows a declination as the MAE decreases. Due to the lack of metal-poor training samples, the neural network performs relatively poorly for stars with [Fe/H] below -0.5 dex. There is a slight increase in prediction uncertainty when [Fe/H] is extended to 0.5 dex, which is also caused by the nonuniformity distribution of training samples. The synchronization between the changes of the prediction uncertainties and the residual shows that the uncertainty given by the neural network is able to be used as a reference for the accuracy of predictions.

\section{Apply to big samples from LAMOST DR8}
\label{sec:4}
In section~\ref{sec:3}, we studied the performance of our neural network on the training and the test sets. We demonstrated that the neural network is capable of making accurate predictions for the vast majority of stars. As for the stars lie in the boundary area of the parameter range, they can be further screened using uncertainties come along with the NN predictions. In this section, we will apply the well-trained network to the large sample of LAMOST DR8 low-resolution spectra and analyse the prediction results.
\subsection{Distribution in chemical spaces}
\begin{figure*}
 \includegraphics[width=\textwidth]{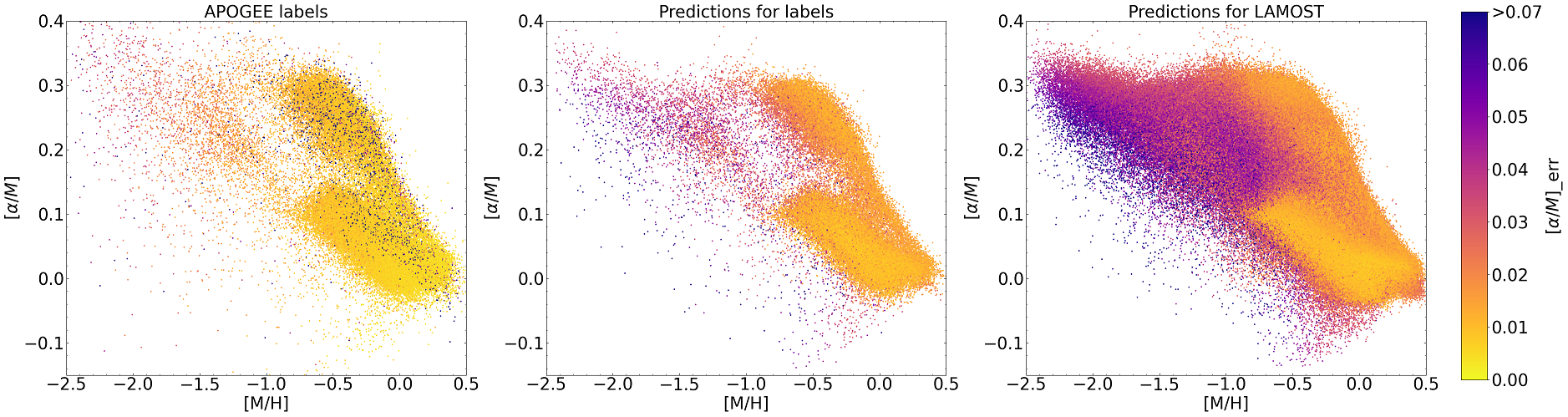}
 \caption{Distribution of samples in [$\alpha$/M] vs. [M/H] plane for APOGEE labels, predictions for the labels and predictions for 1,210,145 giants in LAMOST DR8 catalog. All the subplots are color-scaled by the uncertainty. For the middle and the right subplots, a color closer to yellow indicates that the neural network is more confident in the predicted value, and dots colored close to purple represent predictions that may not be accurate enough.}
 \label{fig:12}
\end{figure*}

In the predictions given by the neural network, there are two values that do not represent the abundance of individual elements, they are the total $\alpha$ abundance [$\alpha$/M] and the total metallicity [M/H]. In Figure~\ref{fig:12}, we show the distribution of samples in [$\alpha$/M] vs. [M/H] plane. The left subplot shows the distribution of the original APOGEE labels, the middle subplot shows the distribution of the predictions for the training and the test sets and the right one shows the situation for LAMOST DR8 samples. The uncertainty of each prediction is indicated by colour, the dots whose colour is closer to yellow are predictions that the neural network is more confident about. From a morphological point of view, the three pictures are very similar, which shows that the predictions of the neural network are accurate. In the middle and the right subplot, the uncertainty of the prediction for stars with low $\alpha$ abundance is relatively higher. Besides, metal-poor stars also tend to have higher uncertainty, which is what we expected. The high uncertainty doesn’t necessarily mean that the prediction is inaccurate, since the prediction is the mean of predictions in the 100 times of forward passes, but it can certainly be a tool that allows users to leave better predictions.\par
\begin{figure}
 \includegraphics[width=\columnwidth]{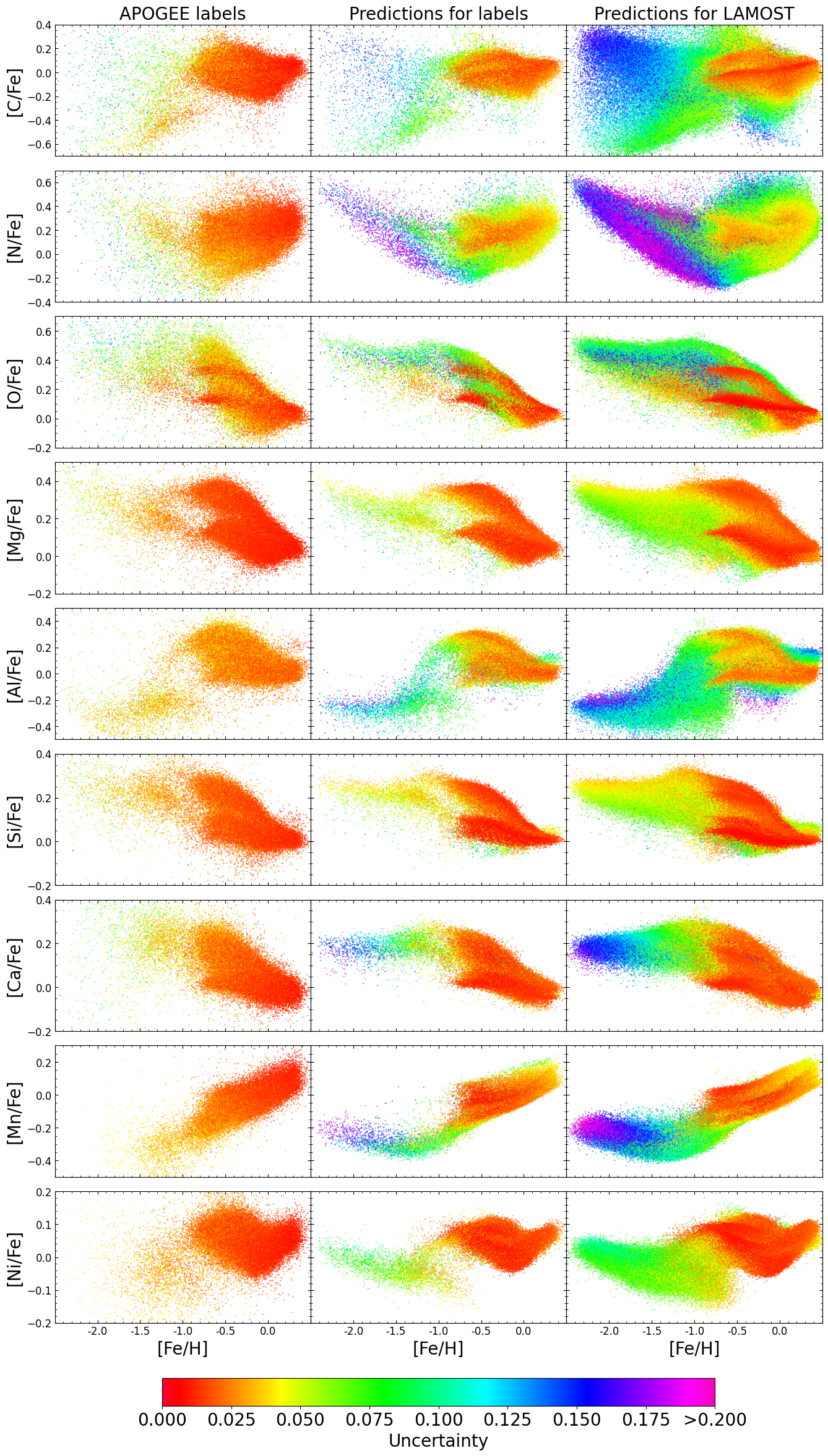}
 \caption{ Distribution of samples in [X/Fe] vs. [Fe/H] planes for APOGEE labels, predictions for the labels and predictions for 1,210,145 giants in LAMOST DR8 catalog. Subplots for each element share the y-axis, and the x-axis represents [Fe/H] that ranges from -2.5 dex to 0.5 dex. Each subplot is color-scaled by the uncertainty. For subplots in the middle and the right columns, dots colored closer to red represent predictions that could be more accurate, a color closer to purple indicates higher uncertainty.}
 \label{fig:13}
\end{figure}

Similar to Figure~\ref{fig:12}, Figure~\ref{fig:13} presents the distribution of samples in [X/Fe] vs. [Fe/H] planes for nine elements. From the left to the right, the subplots show the distribution of APOGEE labels, predictions for samples in the training and the test set and predictions for LAMOST DR8 samples, respectively. Subplots for each element share the y-axis and all the subplots share a color bar. The color of each point in the subplots represents the uncertainty of the APOGEE label or the neural network's prediction for a certain star. In the middle subplots, the neural network not only give predictions for stars with APOGEE labels, but also predict on stars with empty labels, so there are more stars in the middle subplots than the left ones. For manganese, nickel and calcium, the trend for stars with [Fe/H] < -1.5 dex is not clear in the training set because the samples are too sparse. In the middle subplots, the predictions for these stars are usually in a narrow range and with high uncertainty but the tendency seems to be coherent with the whole and the range get wider in the predictions for LAMOST as shown in the right subplots. For silicon in the left subplot, the distribution of stars with [Fe/H] < -2.0 dex shows significant scatter, which can also be seen in \citet[see Figure 1]{shi2009statistical} based on high-resolution spectral analysis. Although the predictions of the neural network are consistent with the main trend in the left subplot, but it can not cover all the diffuse points. In the right subplot of aluminum, there is a plateau in the high aluminum abundance fraction. The same situation does not appear in the middle subplot, which means that it is not a problem with the network structure but the training samples. Most of our samples have aluminum abundances lower than 0.40 dex, the upper bound of our predictions is about 0.35 dex, because of boundary effect. Stars with [Al/Fe] higher than the upper bound will be underestimated, thus forming the plateau. This plateau is an extreme embodiment of the boundary effect of neural networks, it consists of stars with relatively large uncertainty. The phenomena mentioned above also occur in the predictions of other elements, but not as severe as the elements we are discussed. The uncertainty of predictions for stars that lie in the boundary area of the abundance range and metal-poor stars is relatively higher because of the lack of samples in the training set. These predictions should be treated carefully, but they are not necessarily wrong, and can be used as references. The predictions for stars with [Fe/H] > 0.0 dex is reliable, their uncertainties are generally small and the trends for magnesium, aluminum, silicon and calcium are consistent with the trends for the same elements in high-resolution analysis of super metal-rich stars by \citet{chen2003chemical}. For all the elements, the shapes of the three subplots are generally similar, so we consider the predictions of the neural network to be credible.\par
\begin{figure*}
 \includegraphics[width=\textwidth]{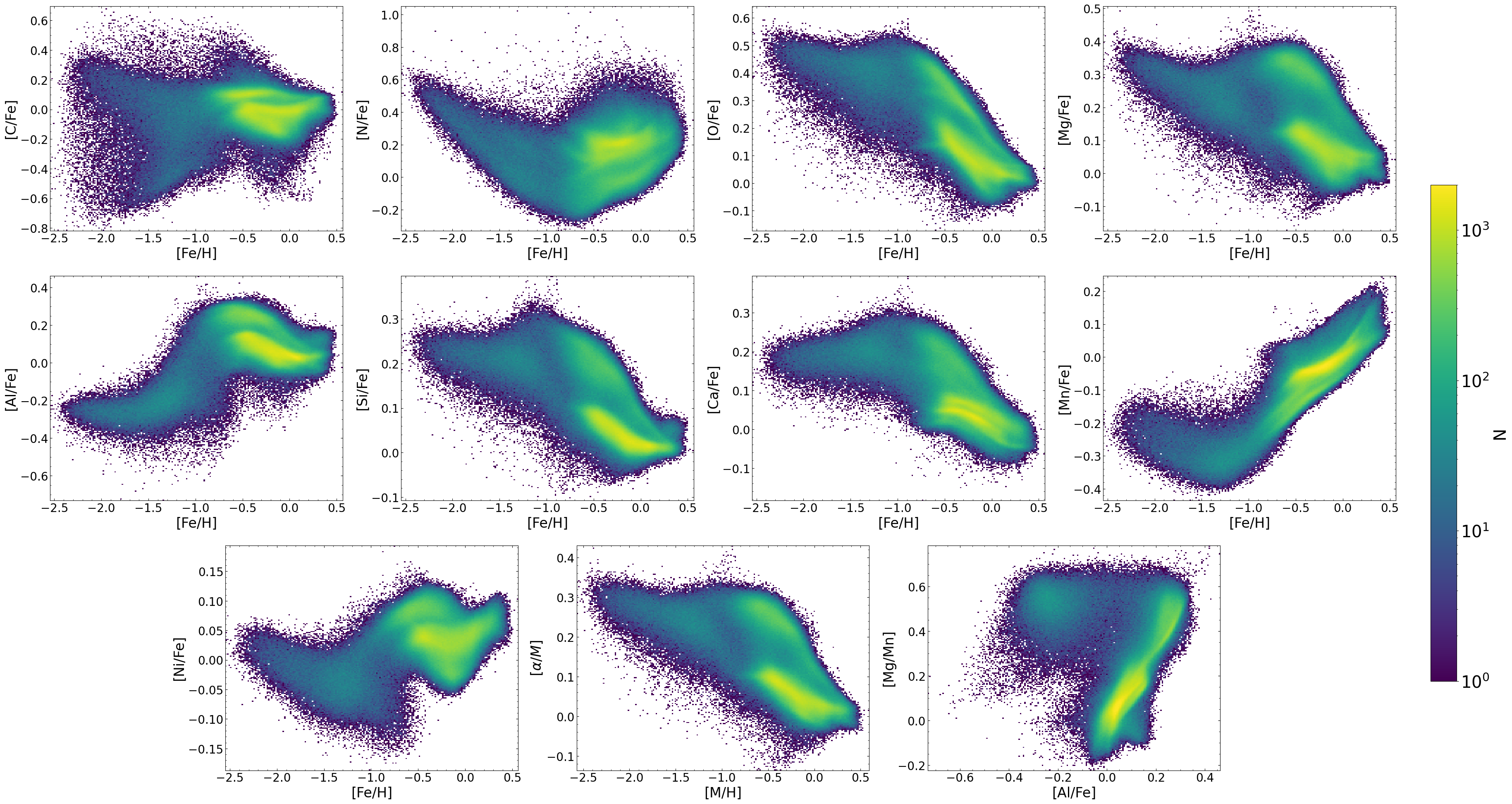}
 \caption{ Distribution of samples from LAMOST DR8 in different chemical planes. The color represents the number of samples in each bin, a brighter color indicates a larger sample size.}
 \label{fig:14}
\end{figure*}

Figure~\ref{fig:14} is a density plot of the distribution of samples from LAMOST DR8 in different chemical planes. We divided both the x-axis and y-axis into 200 bins, the color indicates the number of stars in each bin. We can clearly see the fine aggregated structures in the subplots, especially in the part that [Fe/H] is higher than -1.0 dex. In the bottom right subplot, we present the distribution in the space of [Mg/Mn] vs. [Al/Fe] (e.g., \citealt{das2020ages}) and the different components are distinguishable. The vast majority of our results are consistent with trends in the APOGEE labels, but it is worth emphasizing that for stars with [Fe/H] < -1.5 dex, the predictions of [Mn/Fe], [Ni/Fe], [Ca/Fe] should be treated carefully, because training samples are very sparse or non-existent in these ranges.
\subsection{Independent verification with GALAH DR3}
\begin{figure*}
 \includegraphics[width=\textwidth]{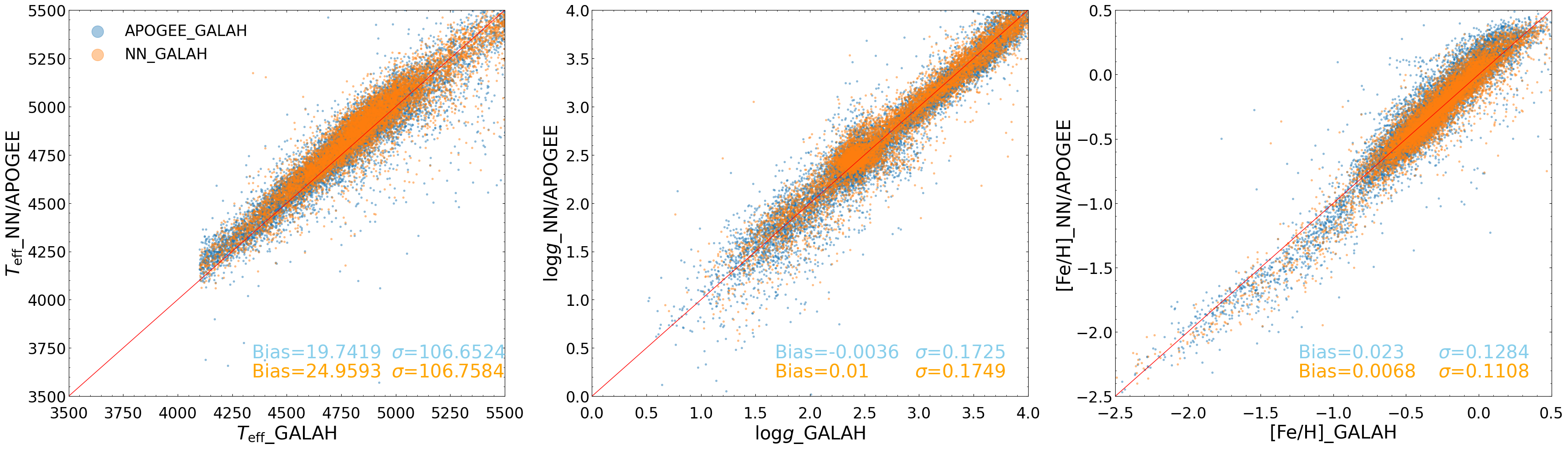}
 \caption{Cross-validation between NN predictions and GALAH DR3. The orange dots represent the comparison between the NN predictions and GALAH, the blue dots represent the comparison between APOGEE and GALAH. The arithmetic mean and standard deviation of the residuals are represented in the subplots as Bias and $\sigma$, respectively.}
 \label{fig:15}
\end{figure*}

To test the accuracy of our results, we use high-resolution for cross-validation. We cross-matched the stars in our results with stars from GALAH DR3 \citep{buder2021galah+} catalog, and obtained 11,949 stars in common. For reference, we also cross-matched APOGEE DR17 catalog with GALAH DR3 catalog, there are 39,403 stars in common, and we selected 15,408 of them with the same stellar parameters range as the training set. Figure~\ref{fig:15} shows the distribution of the three stellar parameters, $T\rm{_{eff}}$, log \textit{g} and [Fe/H]. The x-axis is the value given by GALAH while the y-axis is the predicted value given by the neural network or the APOGEE labels. The orange dots represent the comparison between the NN predictions and GALAH, the blue dots represent the comparison between APOGEE and GALAH. When we study effective temperatures, surface gravity and [Fe/H], we chose stars whose \verb'flag_sp' from GALAH is equal to 0, which results in a cut-off at 4100 K for the effective temperature measured by GALAH. The residuals are defined as neural network predictions minus GALAH labels or APOGEE labels minus GALAH labels. We present the mean value of the residuals as bias and the standard deviation as $\sigma$ in the lower right corner of each subplot, orange represents value for the residual between NN predictions and GALAH labels, blue represents value for the residual between measurements of the two surveys. There seems to be a systematic error between our predictions and the GALAH labels, since the bias is about 25 K. On the other hand, the bias between APOGEE and GALAH labels is 20 K, which means there also is systematic error between the measurements of APOGEE and GALAH. Since our neural network is trained with labels from APOGEE, we consider that the bias between our predictions and GALAH originates from the systematic bias between APOGEE and GALAH, which has been discussed in \citet{nandakumar2022combined}. The bias between the neural network predictions and the GALAH labels are very small on the surface gravity and iron abundance, but our predictions are systematically smaller for stars with [Fe/H] given by GALAH between -1.7 dex and -0.9 dex, and systematically larger for metal-rich stars. This difference also originates from the difference between APOGEE and GALAH, because the blue dots show a same trend in the right subplot. In each subplot, the deviations of the two different common star sets are very close, showing that the neural network performs like APOGEE did.\par
\begin{figure*}
 \includegraphics[width=0.85\textwidth]{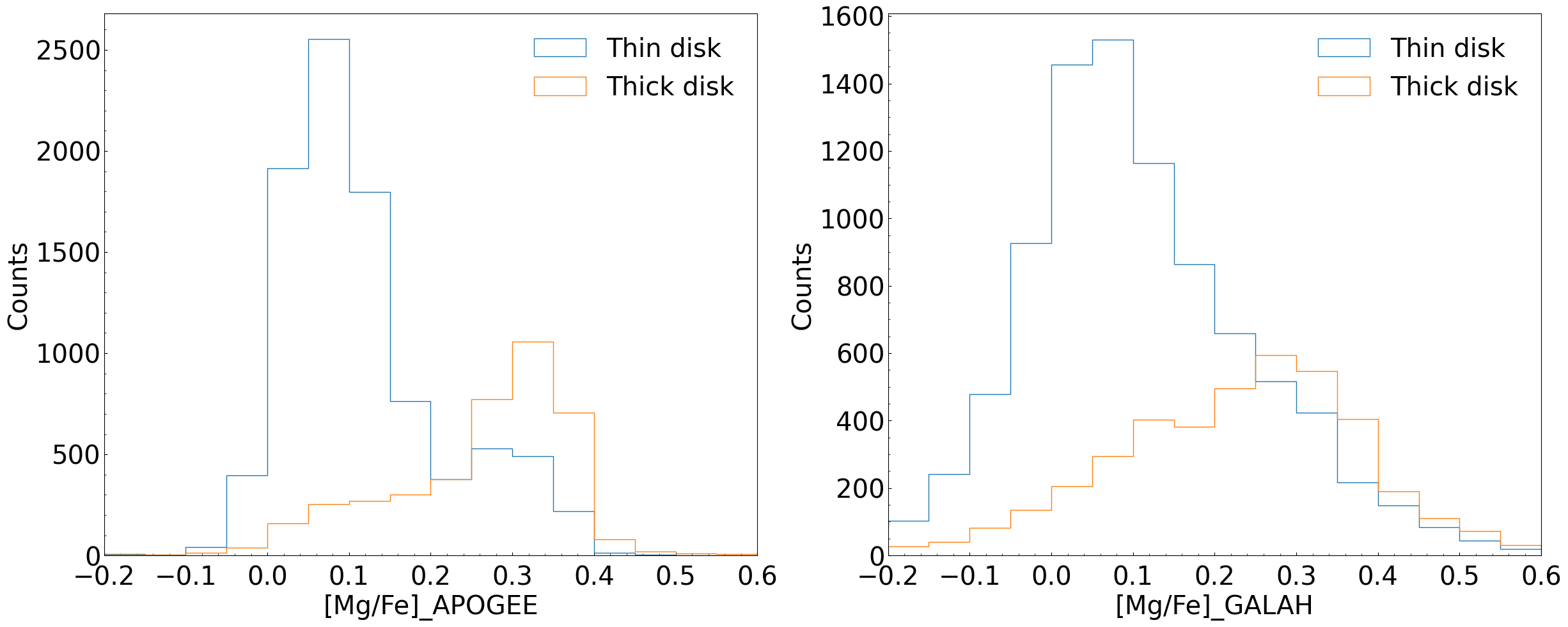}
 \caption{Distribution of [Mg/Fe] measured by APOGEE and GALAH. The blue and orange histograms represent the kinematically selected thin disk and thick disk stars, respectively.}
 \label{fig:16}
\end{figure*}

The GALAH survey obtained spectra in 4 narrow optical bands, while the APOGEE survey obtained spectra in the H-band, there are inevitably some systematic differences between their measurements. We computed velocities in a Galactocentric cylindrical coordinate system for the APOGEE-GALAH common stars and roughly divided them into thick disk stars and thin disk stars through setting the maximum speed of thin disk stars to be 85 km s$^{-1}$ (e.g., \citealt{sitnova2015systematic}). The kinematic parameters including velocities and positions were calculated by Zhang et al. (in preparation, and reference therein). In Figure~\ref{fig:16}, we present the distribution of [Mg/Fe] measured by APOGEE and GALAH as an example. The left subplot shows the distribution of magnesium abundance measured by APOGEE, the thick disk and the thin disk are relatively independent, and there is a demarcation at [Mg/Fe] = 0.2 dex. The magnesium-rich part is dominated by thick disk stars while [Mg/Fe] is generally lower in the thin disk stars. In the right subplot, thick disk stars and thin disk stars seem to be mixed together, and there is no obvious demarcation. We do not apply further comparisons between the neural network predictions and the GALAH DR3 labels for individual elements, because our network is trained with APOGEE labels and similar difference will also appear in the predictions of the neural network. Considering the kinematic results, we prefer the measurements of APOGEE DR17 and the neural network predictions.\par
It is important to note that for some elements, there are no clear features in the low-resolution spectra from LAMOST. In the case where it seems impossible to derive the abundances, the neural network will still provide its results. Unlike \texttt{The Cannon} or \texttt{The Payne}, our neural network builds the mapping directly from spectra to labels, which results in poorer model interpretability and difficulties in explaining the source of these results one-to-one. But judging from the results in the test set, our neural network did a decent job and the abundances could be inferred from their astrophysical correlations, such as the nucleosynthesis reaction network among different elements.

\section{Conclusion}
\label{sec:5}
In this work, we built a deep neural network based on an open-source python package called \texttt{astroNN} to estimate stellar atmospheric parameters and nine elemental abundances for giants from LAMOST DR8. The neural network is trained on common stars of APOGEE DR17 catalog and LAMOST DR8 catalog, using spectra from LAMOST and labels from APOGEE. On the test set, the mean absolute error is 29 K for effective temperature, 0.07 dex for surface gravity, 0.03 dex for [Fe/H] and [M/H] and 0.02 dex for [$\alpha$/M]. Typical MAE range for the nine elements is 0.02 dex to 0.04 dex. The residuals between the APOGEE labels and neural network predictions are mostly close to zero for both training and test sets. Statistically, the neural network performs well and we consider it to have the ability to derive multiple abundances simultaneously and accurately from low-resolution spectra.\par
The trained neural network was than applied to 1,210,145 low-resolution spectra from LAMOST DR8, the distribution of the predictions in [X/Fe] vs. [Fe/H] planes is similar to that of the APOGEE labels and the predictions for training samples. The mean uncertainty is 39 K for effective temperature, 0.10 dex for log \textit{g}, 0.04 dex for [Fe/H] and [M/H] and 0.02 dex for [$\alpha$/M]. For nitrogen and oxygen, the mean uncertainty is 0.05 dex. The mean uncertainty for other elements is in the range of 0.02 dex to 0.04 dex. After we got the predictions for LAMOST DR8 spectra, we cross-matched our results with GALAH DR3 catalog to test the accuracy of the neural network’s predictions of $T\rm{_{eff}}$, log \textit{g} and [Fe/H]. It turns out that the consistency is generally good, but there are some systematic errors that originates from the systematic bias between APOGEE and GALAH. The systematic bias between the two surveys is more pronounced in elemental abundance measurements. Based on the kinematically-selected thin and thick disk stars, we prefer the measurements of APOGEE and the predictions given by the neural network trained with APOGEE labels.\par
We have compiled all of our results into a value added catalog (VAC), which contains atmospheric parameters and individual elemental abundances for 1.2 million giants in the latest data release of LAMOST. A large sample of stars with detailed chemical abundance information is of great help to our understanding of the formation and evolution of the Milky Way, the VAC makes up the lack of abundance information in the LAMOST low-resolution spectroscopic survey and opens up more possibilities for researches based on LAMOST data.
\par
In addition to the neural network highlighted in this article, we have trained a number of sub-networks. Each sub-network estimates the abundance of a specific element along with $T\rm{_{eff}}$, log \textit{g} and [Fe/H]. The predictions given by these sub-networks can be used as a supplement to the predictions of the main network. This study focused on introducing our method and presenting the results on giants. The results on dwarfs will be presented in the future and we will combine the newly released data from Gaia to study substructures in the Milky Way from both chemical and kinematical space. We are also going to take attempt in transfer learning and try to transfer the neural network discussed in this study to other surveys to obtain more informations on the elemental abundances of giants.

\section*{Acknowledgements}
\addcontentsline{toc}{section}{Acknowledgements}

This study is supported by the National Natural Science Foundation of China under grant Nos. 11988101, 11890694, 11973048 and National Key R\&D Program of China No. 2019YFA0405500. ZL thanks Drs. Haining Li, Lan Zhang, Xiangxiang Xue, Wenbo Wu and Mr. Haopeng Zhang for their discussion and useful suggestions. This work is based on data acquired through the Guoshoujing Telescope. Guoshoujing Telescope (the Large Sky Area Multi-Object Fiber Spectroscopic Telescope LAMOST) is a National Major Scientific Project built by the Chinese Academy of Sciences. Funding for the project has been provided by the National Development and Reform Commission. LAMOST is operated and managed by the National Astronomical Observatories, Chinese Academy of Sciences. \par
Funding for the Sloan Digital Sky 
Survey IV has been provided by the 
Alfred P. Sloan Foundation, the U.S. 
Department of Energy Office of 
Science, and the Participating 
Institutions.
SDSS-IV acknowledges support and 
resources from the Center for High 
Performance Computing  at the 
University of Utah. The SDSS 
website is www.sdss.org.

SDSS-IV is managed by the 
Astrophysical Research Consortium 
for the Participating Institutions 
of the SDSS Collaboration including 
the Brazilian Participation Group, 
the Carnegie Institution for Science, 
Carnegie Mellon University, Center for 
Astrophysics | Harvard \& 
Smithsonian, the Chilean Participation 
Group, the French Participation Group, 
Instituto de Astrof\'isica de 
Canarias, The Johns Hopkins 
University, Kavli Institute for the 
Physics and Mathematics of the 
Universe (IPMU) / University of 
Tokyo, the Korean Participation Group, 
Lawrence Berkeley National Laboratory, 
Leibniz Institut f\"ur Astrophysik 
Potsdam (AIP),  Max-Planck-Institut 
f\"ur Astronomie (MPIA Heidelberg), 
Max-Planck-Institut f\"ur 
Astrophysik (MPA Garching), 
Max-Planck-Institut f\"ur 
Extraterrestrische Physik (MPE), 
National Astronomical Observatories of 
China, New Mexico State University, 
New York University, University of 
Notre Dame, Observat\'ario 
Nacional / MCTI, The Ohio State 
University, Pennsylvania State 
University, Shanghai 
Astronomical Observatory, United 
Kingdom Participation Group, 
Universidad Nacional Aut\'onoma 
de M\'exico, University of Arizona, 
University of Colorado Boulder, 
University of Oxford, University of 
Portsmouth, University of Utah, 
University of Virginia, University 
of Washington, University of 
Wisconsin, Vanderbilt University, 
and Yale University. \par
This work made use of the Third Data Release of the GALAH Survey \citep{buder2021galah+}. The GALAH Survey is based on data acquired through the Australian Astronomical Observatory, under programs: A/2013B/13 (The GALAH pilot survey); A/2014A/25, A/2015A/19, A2017A/18 (The GALAH survey phase 1); A2018A/18 (Open clusters with HERMES); A2019A/1 (Hierarchical star formation in Ori OB1); A2019A/15 (The GALAH survey phase 2); A/2015B/19, A/2016A/22, A/2016B/10, A/2017B/16, A/2018B/15 (The HERMES-TESS program); and A/2015A/3, A/2015B/1, A/2015B/19, A/2016A/22, A/2016B/12, A/2017A/14 (The HERMES K2-follow-up program). We acknowledge the traditional owners of the land on which the AAT stands, the Gamilaraay people, and pay our respects to elders past and present. This paper includes data that has been provided by AAO Data Central (datacentral.org.au).

\section*{Data Availability}
The VAC is available at \url{http://www.lamost.org/dr8/v1.1/doc/vac}.



\bibliographystyle{mnras}
\bibliography{References} 

\begin{thebibliography}{}
\makeatletter
\relax
\def\mn@urlcharsother{\let\do\@makeother \do\$\do\&\do\#\do\^\do\_\do\%\do\~}
\def\mn@doi{\begingroup\mn@urlcharsother \@ifnextchar [ {\mn@doi@}
  {\mn@doi@[]}}
\def\mn@doi@[#1]#2{\def\@tempa{#1}\ifx\@tempa\@empty \href
  {http://dx.doi.org/#2} {doi:#2}\else \href {http://dx.doi.org/#2} {#1}\fi
  \endgroup}
\def\mn@eprint#1#2{\mn@eprint@#1:#2::\@nil}
\def\mn@eprint@arXiv#1{\href {http://arxiv.org/abs/#1} {{\tt arXiv:#1}}}
\def\mn@eprint@dblp#1{\href {http://dblp.uni-trier.de/rec/bibtex/#1.xml}
  {dblp:#1}}
\def\mn@eprint@#1:#2:#3:#4\@nil{\def\@tempa {#1}\def\@tempb {#2}\def\@tempc
  {#3}\ifx \@tempc \@empty \let \@tempc \@tempb \let \@tempb \@tempa \fi \ifx
  \@tempb \@empty \def\@tempb {arXiv}\fi \@ifundefined
  {mn@eprint@\@tempb}{\@tempb:\@tempc}{\expandafter \expandafter \csname
  mn@eprint@\@tempb\endcsname \expandafter{\@tempc}}}

\bibitem[\protect\citeauthoryear{Abolfathi et~al.,}{Abolfathi
  et~al.}{2018}]{abolfathi2018fourteenth}
Abolfathi B.,  et~al., 2018, The Astrophysical Journal Supplement Series, 235,
  42

\bibitem[\protect\citeauthoryear{Accetta et~al.,}{Accetta
  et~al.}{2022}]{accetta2022seventeenth}
Accetta K.,  et~al., 2022, The Astrophysical Journal Supplement Series, 259, 35

\bibitem[\protect\citeauthoryear{Bailer-Jones, Irwin, Gilmore  \& von
  Hippel}{Bailer-Jones et~al.}{1997}]{bailer1997physical}
Bailer-Jones C.~A.,  Irwin M.,  Gilmore G.,   von Hippel T.,  1997, Monthly
  Notices of the Royal Astronomical Society, 292, 157

\bibitem[\protect\citeauthoryear{Belokurov, Erkal, Evans, Koposov  \&
  Deason}{Belokurov et~al.}{2018}]{belokurov2018co}
Belokurov V.,  Erkal D.,  Evans N.,  Koposov S.,   Deason A.,  2018, Monthly
  Notices of the Royal Astronomical Society, 478, 611

\bibitem[\protect\citeauthoryear{Buder et~al.,}{Buder
  et~al.}{2021}]{buder2021galah+}
Buder S.,  et~al., 2021, Monthly Notices of the Royal Astronomical Society,
  506, 150

\bibitem[\protect\citeauthoryear{Chen, Zhao, Nissen, Bai  \& Qiu}{Chen
  et~al.}{2003}]{chen2003chemical}
Chen Y.,  Zhao G.,  Nissen P.,  Bai G.,   Qiu H.,  2003, The Astrophysical
  Journal, 591, 925

\bibitem[\protect\citeauthoryear{Cui et~al.,}{Cui et~al.}{2012}]{cui2012large}
Cui X.-Q.,  et~al., 2012, Research in Astronomy and Astrophysics, 12, 1197

\bibitem[\protect\citeauthoryear{Damianou \& Lawrence}{Damianou \&
  Lawrence}{2013}]{damianou2013deep}
Damianou A.,  Lawrence N.~D.,  2013, in Artificial intelligence and statistics.
  pp 207--215

\bibitem[\protect\citeauthoryear{Das, Hawkins  \& Jofr{\'e}}{Das
  et~al.}{2020}]{das2020ages}
Das P.,  Hawkins K.,   Jofr{\'e} P.,  2020, Monthly Notices of the Royal
  Astronomical Society, 493, 5195

\bibitem[\protect\citeauthoryear{De~Silva et~al.,}{De~Silva
  et~al.}{2015}]{de2015galah}
De~Silva G.~M.,  et~al., 2015, Monthly Notices of the Royal Astronomical
  Society, 449, 2604

\bibitem[\protect\citeauthoryear{Deng et~al.,}{Deng
  et~al.}{2012}]{deng2012lamost}
Deng L.-C.,  et~al., 2012, Research in Astronomy and Astrophysics, 12, 735

\bibitem[\protect\citeauthoryear{Fabbro, Venn, O'Briain, Bialek, Kielty,
  Jahandar  \& Monty}{Fabbro et~al.}{2018}]{fabbro2018application}
Fabbro S.,  Venn K.,  O'Briain T.,  Bialek S.,  Kielty C.,  Jahandar F.,
  Monty S.,  2018, Monthly Notices of the Royal Astronomical Society, 475, 2978

\bibitem[\protect\citeauthoryear{Feuillet, Sahlholdt, Feltzing  \&
  Casagrande}{Feuillet et~al.}{2021}]{feuillet2021selecting}
Feuillet D.~K.,  Sahlholdt C.~L.,  Feltzing S.,   Casagrande L.,  2021, Monthly
  Notices of the Royal Astronomical Society, 508, 1489

\bibitem[\protect\citeauthoryear{Gal \& Ghahramani}{Gal \&
  Ghahramani}{2016}]{gal2016dropout}
Gal Y.,  Ghahramani Z.,  2016, in international conference on machine learning.
  pp 1050--1059

\bibitem[\protect\citeauthoryear{Gilmore et~al.,}{Gilmore
  et~al.}{2012}]{gilmore2012gaia}
Gilmore G.,  et~al., 2012, Messenger, 147, 25

\bibitem[\protect\citeauthoryear{Helmi}{Helmi}{2020}]{helmi2020streams}
Helmi A.,  2020, Annual Review of Astronomy and Astrophysics, 58, 205

\bibitem[\protect\citeauthoryear{Helmi, White, De~Zeeuw  \& Zhao}{Helmi
  et~al.}{1999}]{helmi1999debris}
Helmi A.,  White S.~D.,  De~Zeeuw P.~T.,   Zhao H.,  1999, Nature, 402, 53

\bibitem[\protect\citeauthoryear{Helmi, Babusiaux, Koppelman, Massari,
  Veljanoski  \& Brown}{Helmi et~al.}{2018}]{helmi2018merger}
Helmi A.,  Babusiaux C.,  Koppelman H.~H.,  Massari D.,  Veljanoski J.,   Brown
  A.~G.,  2018, Nature, 563, 85

\bibitem[\protect\citeauthoryear{Hinton, Srivastava, Krizhevsky, Sutskever  \&
  Salakhutdinov}{Hinton et~al.}{2012}]{hinton2012improving}
Hinton G.~E.,  Srivastava N.,  Krizhevsky A.,  Sutskever I.,   Salakhutdinov
  R.~R.,  2012, arXiv preprint arXiv:1207.0580

\bibitem[\protect\citeauthoryear{Ho, Ness, Hogg  \& Rix}{Ho
  et~al.}{2016}]{ho2016cannon}
Ho A.~Y.,  Ness M.,  Hogg D.~W.,   Rix H.-W.,  2016, Astrophysics Source Code
  Library, pp ascl--1602

\bibitem[\protect\citeauthoryear{Ho et~al.,}{Ho et~al.}{2017}]{ho2017label}
Ho A.~Y.,  et~al., 2017, The Astrophysical Journal, 836, 5

\bibitem[\protect\citeauthoryear{Imig et~al.,}{Imig
  et~al.}{2022}]{imig2022sdss}
Imig J.,  et~al., 2022, The Astronomical Journal, 163, 56

\bibitem[\protect\citeauthoryear{J{\"o}nsson et~al.,}{J{\"o}nsson
  et~al.}{2020}]{jonsson2020apogee}
J{\"o}nsson H.,  et~al., 2020, The Astronomical Journal, 160, 120

\bibitem[\protect\citeauthoryear{Kendall \& Gal}{Kendall \&
  Gal}{2017}]{kendall2017uncertainties}
Kendall A.,  Gal Y.,  2017, Advances in neural information processing systems,
  30

\bibitem[\protect\citeauthoryear{Koppelman, Helmi, Massari, Price-Whelan  \&
  Starkenburg}{Koppelman et~al.}{2019}]{koppelman2019multiple}
Koppelman H.~H.,  Helmi A.,  Massari D.,  Price-Whelan A.~M.,   Starkenburg
  T.~K.,  2019, Astronomy \& Astrophysics, 631, L9

\bibitem[\protect\citeauthoryear{LeCun, Bengio  \& Hinton}{LeCun
  et~al.}{2015}]{lecun2015deep}
LeCun Y.,  Bengio Y.,   Hinton G.,  2015, nature, 521, 436

\bibitem[\protect\citeauthoryear{Leung \& Bovy}{Leung \&
  Bovy}{2019}]{leung2019deep}
Leung H.~W.,  Bovy J.,  2019, Monthly Notices of the Royal Astronomical
  Society, 483, 3255

\bibitem[\protect\citeauthoryear{Li, Zhao, Christlieb, Wang, Wang, Zhang, Hou
  \& Yuan}{Li et~al.}{2015}]{li2015spectroscopic}
Li H.-N.,  Zhao G.,  Christlieb N.,  Wang L.,  Wang W.,  Zhang Y.,  Hou Y.,
  Yuan H.,  2015, The Astrophysical Journal, 798, 110

\bibitem[\protect\citeauthoryear{Liang, Zhao, Chen, Zuo, Zhang, Zhu  \&
  Zhao}{Liang et~al.}{2019}]{liang2019elemental}
Liang X.,  Zhao J.,  Chen Y.,  Zuo W.,  Zhang J.,  Zhu J.,   Zhao G.,  2019,
  The Astrophysical Journal, 887, 193

\bibitem[\protect\citeauthoryear{Liu et~al.,}{Liu et~al.}{2013}]{liu2013lss}
Liu X.-W.,  et~al., 2013, Proceedings of the International Astronomical Union,
  9, 310

\bibitem[\protect\citeauthoryear{Liu, Zhao  \& Hou}{Liu
  et~al.}{2015}]{liu2015preface}
Liu X.-W.,  Zhao G.,   Hou J.-L.,  2015, Research in Astronomy and
  Astrophysics, 15, 1089

\bibitem[\protect\citeauthoryear{Luo et~al.,}{Luo et~al.}{2015}]{luo2015first}
Luo A.-L.,  et~al., 2015, Research in Astronomy and Astrophysics, 15, 1095

\bibitem[\protect\citeauthoryear{Majewski et~al.,}{Majewski
  et~al.}{2017}]{majewski2017apache}
Majewski S.~R.,  et~al., 2017, The Astronomical Journal, 154, 94

\bibitem[\protect\citeauthoryear{Myeong, Vasiliev, Iorio, Evans  \&
  Belokurov}{Myeong et~al.}{2019}]{myeong2019evidence}
Myeong G.,  Vasiliev E.,  Iorio G.,  Evans N.,   Belokurov V.,  2019, Monthly
  Notices of the Royal Astronomical Society, 488, 1235

\bibitem[\protect\citeauthoryear{Nandakumar et~al.,}{Nandakumar
  et~al.}{2022}]{nandakumar2022combined}
Nandakumar G.,  et~al., 2022, Monthly Notices of the Royal Astronomical
  Society, 513, 232

\bibitem[\protect\citeauthoryear{Ness, Hogg, Rix, Ho  \& Zasowski}{Ness
  et~al.}{2015}]{ness2015cannon}
Ness M.,  Hogg D.~W.,  Rix H.-W.,  Ho A.~Y.,   Zasowski G.,  2015, The
  Astrophysical Journal, 808, 16

\bibitem[\protect\citeauthoryear{Prusti et~al.,}{Prusti
  et~al.}{2016}]{prusti2016gaia}
Prusti T.,  et~al., 2016, Astronomy \& astrophysics, 595, A1

\bibitem[\protect\citeauthoryear{Shi, Gehren, Mashonkina  \& Zhao}{Shi
  et~al.}{2009}]{shi2009statistical}
Shi J.,  Gehren T.,  Mashonkina L.,   Zhao G.,  2009, Astronomy \&
  Astrophysics, 503, 533

\bibitem[\protect\citeauthoryear{Sitnova et~al.,}{Sitnova
  et~al.}{2015}]{sitnova2015systematic}
Sitnova T.,  et~al., 2015, The Astrophysical Journal, 808, 148

\bibitem[\protect\citeauthoryear{Steinmetz et~al.,}{Steinmetz
  et~al.}{2006}]{steinmetz2006radial}
Steinmetz M.,  et~al., 2006, The Astronomical Journal, 132, 1645

\bibitem[\protect\citeauthoryear{Ting, Rix, Conroy, Ho  \& Lin}{Ting
  et~al.}{2017}]{ting2017measuring}
Ting Y.-S.,  Rix H.-W.,  Conroy C.,  Ho A.~Y.,   Lin J.,  2017, The
  Astrophysical Journal Letters, 849, L9

\bibitem[\protect\citeauthoryear{Ting, Conroy, Rix  \& Cargile}{Ting
  et~al.}{2019}]{ting2019payne}
Ting Y.-S.,  Conroy C.,  Rix H.-W.,   Cargile P.,  2019, The Astrophysical
  Journal, 879, 69

\bibitem[\protect\citeauthoryear{Von~Hippel, Storrie-Lombardi, Storrie-Lombardi
   \& Irwin}{Von~Hippel et~al.}{1994}]{von1994automated}
Von~Hippel T.,  Storrie-Lombardi L.,  Storrie-Lombardi M.,   Irwin M.,  1994,
  Monthly Notices of the Royal Astronomical Society, 269, 97

\bibitem[\protect\citeauthoryear{Xiang et~al.,}{Xiang
  et~al.}{2019}]{xiang2019abundance}
Xiang M.,  et~al., 2019, The Astrophysical Journal Supplement Series, 245, 34

\bibitem[\protect\citeauthoryear{Yan et~al.,}{Yan et~al.}{2019}]{yan2019sdss}
Yan R.,  et~al., 2019, The Astrophysical Journal, 883, 175

\bibitem[\protect\citeauthoryear{Yanny et~al.,}{Yanny
  et~al.}{2009}]{yanny2009segue}
Yanny B.,  et~al., 2009, The Astronomical Journal, 137, 4377

\bibitem[\protect\citeauthoryear{Zhang, Zhao, Yang, Wang  \& Zuo}{Zhang
  et~al.}{2019}]{zhang2019938}
Zhang X.,  Zhao G.,  Yang C.,  Wang Q.,   Zuo W.,  2019, Publications of the
  Astronomical Society of the Pacific, 131, 094202

\bibitem[\protect\citeauthoryear{Zhao \& Li}{Zhao \& Li}{2001}]{zhao2001coude}
Zhao G.,  Li H.-B.,  2001, Chinese Journal of Astronomy and Astrophysics, 1,
  555

\bibitem[\protect\citeauthoryear{Zhao, Qiu  \& Mao}{Zhao
  et~al.}{2001}]{zhao2001high}
Zhao G.,  Qiu H.,   Mao S.,  2001, The Astrophysical Journal, 551, L85

\bibitem[\protect\citeauthoryear{Zhao, Chen, Shi, Liang, Hou, Chen, Zhang  \&
  Li}{Zhao et~al.}{2006}]{zhao2006stellar}
Zhao G.,  Chen Y.-Q.,  Shi J.-R.,  Liang Y.-C.,  Hou J.-L.,  Chen L.,  Zhang
  H.-W.,   Li A.-G.,  2006, Chinese Journal of Astronomy and Astrophysics, 6,
  265

\bibitem[\protect\citeauthoryear{Zhao, Zhao  \& Chen}{Zhao
  et~al.}{2009}]{zhao2009catalog}
Zhao J.,  Zhao G.,   Chen Y.,  2009, The Astrophysical Journal, 692, L113

\bibitem[\protect\citeauthoryear{Zhao, Zhao, Chu, Jing  \& Deng}{Zhao
  et~al.}{2012}]{zhao2012lamost}
Zhao G.,  Zhao Y.-H.,  Chu Y.-Q.,  Jing Y.-P.,   Deng L.-C.,  2012, Research in
  Astronomy and Astrophysics, 12, 723

\makeatother
\end{thebibliography}



\bsp	
\label{lastpage}
\end{document}